\documentclass[onecolumn,noshowpacs,nofootinbib,11pt]{revtex4}
\usepackage{float,xcolor,upgreek,yfonts,tikz}
\usepackage{color} 
\usepackage{graphicx,bigints}
\usepackage{graphicx,float}\usepackage[all]{xy}
\usepackage{amsmath,upgreek,accents}

\usepackage{amssymb}

\usepackage{alphabeta}
\usepackage{textgreek}
\newcommand{\beq}{\begin{eqnarray}}
\newcommand{\eeq}{\end{eqnarray}}
\newcommand{\be}{\begin{equation}}
\newcommand{\ee}{\end{equation}}

\newcommand{\bec}{\begin{center}}
\newcommand{\eec}{\end{center}}
\newcommand{\bea}{\begin{eqnarray}}
\newcommand{\eea}{\end{eqnarray}}

\usepackage{mathrsfs}
\newcommand{\ba}{\begin{eqnarray}}
\newcommand{\ea}{\end{eqnarray}}
\newcommand{\clt}{\textcolor{black}}

\DeclareFontFamily{OT1}{rsfs}{}
\DeclareFontShape{OT1}{rsfs}{m}{n}{ <-7> rsfs5 <7-10> rsfs7 <10->
rsfs10}{} \DeclareMathAlphabet{\mycal}{OT1}{rsfs}{m}{n}

\newcommand{\bes}{\begin{subequations}}
\newcommand{\ees}{\end{subequations}}
\def\ben{\begin{eqnarray}}
\def\een{\end{eqnarray}}
\def\be{\begin{equation}}
\def\ee{\end{equation}}

\usepackage{graphicx}
\usepackage[T3,T1]{fontenc}
\DeclareSymbolFont{tipa}{T3}{cmr}{m}{n}
\DeclareMathAccent{\invbreve}{\mathalpha}{tipa}{16}

\usepackage[colorlinks,hyperindex,unicode]{hyperref}
\definecolor{green1}{RGB}{0,128,0} 
\hypersetup{hidelinks,backref=true,pagebackref=true,hyperindex=true,colorlinks=true,breaklinks=true,urlcolor= blue}
\hypersetup{%
  colorlinks = true,
  linkcolor  = blue,
  citecolor = cyan,
}
\usepackage{bookmark,textgreek}
\usepackage{hyperref,color,xcolor}
\hypersetup{hidelinks,hyperindex=true,colorlinks=true,breaklinks=true,urlcolor= blue}
\hypersetup{%
  colorlinks = true,
  linkcolor  = blue
}

\newcommand\orcidroldao{{\href{https://orcid.org/0000-0003-3978-532X}{\orcidicon}}}
\newcommand\orcidwillians{{\href{https://orcid.org/0000-0001-9750-2637}{\orcidicon}}}
\newcommand\orcidaha{{\href{https://orcid.org/0000-0003-4918-2231}{\orcidicon}}}
\newcommand{\orcidicon}{%
	\begin{tikzpicture}
	\draw[lime, fill=lime] (0,0)
		circle [radius=0.16]
		node[white] {{\fontfamily{qag}\selectfont \tiny ID}};
	\draw[white, fill=white] (-0.0625,0.095)
		circle [radius=0.007];
	\end{tikzpicture}	\hspace{-2mm}
}

\begin{document}

\title{Configurational entropy of generalized sine-Gordon-type models}
\author{W. Barreto\orcidwillians\!\!}
\affiliation{Federal University of ABC, Center of Natural Sciences, Santo Andr\'e, 09580-210, Brazil}\email{willians.barreto@ufabc.edu.br}
\affiliation{Centro de F\'{i}sica Fundamental, Universidad de Los Andes, M\'{e}rida 5101, Venezuela}
\author{A. Herrera--Aguilar\orcidaha\!\!}
\affiliation{Instituto de F\'{i}sica, Benem\'erita Universidad Aut\'onoma de Puebla, Apartado Postal J-48, Puebla, 72570, Mexico.}
\author{R. da Rocha\orcidroldao\!\!}
\affiliation{Federal University of ABC, Center of Mathematics, Santo Andr\'e, 09210-580, Brazil.}
\email{roldao.rocha@ufabc.edu.br}

\begin{abstract}
A family of deformed models of the sine-Gordon-type can be generated by twisting the sine-Gordon model. As  
a particular case, the 3-sine-Gordon model is here addressed, whose differential configurational entropy \clt{and the differential configurational complexity} of three topological sectors are  discussed using two complementary approaches. Stability aspects are also discussed.
\end{abstract}

\maketitle



\section{Introduction}

Configurational entropy (CE) is a measure of information entropy that designates the compression of data and the coding process by any source. The CE quantifies the portion of information encoded into probability distributions underlying the description of physical systems \cite{gs12a,gs12b}. The CE reckons at the Gibbs--Boltzmann entropy, addressing the measure of disorder in (discrete) physical systems. The analog formulation for continuum systems can be implemented by the differential configurational entropy (DCE) \cite{Gleiser:2018kbq,Gleiser:2018jpd,Stephens:2019tav,Bernardini:2016hvx}, whose critical points specify the precise microstates, in the momentum space, which have the higher configurational stability, also designating the most dominant wave state(s) occupied \cite{Gleiser:2015aav,PRDgleiser-graham}. The DCE is a pliable tool for recognizing pattern formation and has been utilized for analyzing a sort of physical systems, including the Hawking--Page transition regulated by a critical configurational instability \cite{Braga:2019jqg,Lee:2021rag}. Refs. \cite{Braga:2020opg,Braga:2017fsb,Braga:2020hhs,Braga:2020myi} employed the DCE to 
study bottomonia and charmonia in AdS/QCD, whose stability in a plasma within the influence of magnetic fields was scrutinized in Ref. \cite{Braga:2021fey}. Other aspects of the DCE in AdS/QCD were revealed in Refs. \cite{Braga:2021zyi}. 
Phase transitions in this context were shown to be promoted by the DCE \cite{Lee:2017ero}. 
A variety of stellar distributions were explored in the DCE setup, \cite{Fernandes-Silva:2019fez,Gleiser:2013mga,Gleiser:2015rwa}. 
 Using the DCE, the mass spectra of hadronic and baryonic states were computed in AdS/QCD models \cite{Bernardini:2018uuy,Braga:2018fyc,daRocha:2021ntm,Barbosa-Cendejas:2018mng,Ferreira:2020iry,Colangelo:2018mrt,Karapetyan:2020yhs,k2,k4,Karapetyan:2018yhm,Karapetyan:2018oye,Karapetyan:2021vyh,Karapetyan:2021crv,Karapetyan:2017edu,Alves:2020cmr}. 
Besides, the DCE addressed (topological) field theories \cite{Bazeia:2018uyg,Bazeia:2021stz}. One can use the DCE to take into
account the underlying information aspects regarding physical models presenting localized energy configurations \cite{Thakur:2020sko}. The DCE approach can be also used to scrutinize 
several non-linear scalar field models and to circumvent intricacies with degenerate energy configurations, selecting 
the best configuration from the informational point of view \cite{Gleiser:2020zaj}. 
 Turbulent aspects of the gravitational collapse massless scalar field in asymptotically AdS spacetime were studied under the DCE apparatus in Refs. \cite{Barreto:2022ohl,Barreto:2022len}.

On investigating the dynamics of some scalar fields, topological solutions emerge, encompassing domain walls and kinks, with applications in several ramifications of nonlinear systems, from high-energy physics to condensed matter physics. Topological defects can describe
phase transitions in the early universe, mapping interfaces separating distinct
regions and also contributing to pattern formation \cite{Dunne,spain}. A topological defect sets in whenever boundary conditions of equations of motion that govern a physical system yield homotopically distinct solutions. When the boundary has a non-trivial homotopy group, which is preserved in differential equations, their solutions are therefore topologically distinct, being classified by their homotopy class. Topological defects are stable not just for small perturbations, however, they can neither decay nor disentangle, since there do not exist continuous mappings between them and the trivial solution.
Topological defects can be evinced in relativistic models described by a
real scalar field $\upphi$ in $(1,1)$-spacetime dimensions, also within the context of deformation
procedures \cite{Epl1,Bazeia:2002xg}. In particular, sine-Gordon models and generalizations have been playing a relevant role in field theory, with remarkable applications 
\cite{Bazeia:2017cqv,epjc,Bazeia:2011ff,sine1,Bernardini:2014mma,Bazeia:2013usa,Bazeia:2004hv,Cruz:2018qby,Gauy:2022xsc,Cruz:2015nrd}. In particular, the seminal work \cite{Epl1}
 instructed how to deform the sine-Gordon model, employing two 
parameters, one of them permitting to set out the specific member in the family of
models, also providing new topological solutions of the field equations. Subsequently, one can derive the energy density associated with the deformed models, corresponding to distinct topological sectors. 
Deformed sine-Gordon-type models can be applied to the study of stable braneworld
models \cite{sine,G}.  
One can scrutinize gravitons on braneworld scenarios, with an extra dimension, that are dynamically generated by topological defects supporting kink
solutions \cite{German:2013jcp}. Therefore scalar fields coupled to gravity regulated by the Einstein--Klein--Gordon system yield thick braneworld scenarios \cite{Barbosa:2014grg}. Moreover, cosmological thick braneworlds that render a natural isotropization mechanism for the early universe \cite{Gog2013prd} as well as braneworlds that generate the expansion of our Universe from extra dimensions \cite{Gog2013plb} are also supported by scalar fields. Besides, bounce-type solutions can be derived in the context of sine-Gordon potentials \cite{Cruz:2014eca}. 

The main aim of this work is to study the DCE \clt{and the differential configurational complexity (DCC)} underlying deformed sine-Gordon-type models, particularly focusing on the 3-sine-Gordon deformed models, which have several analytical approaches. \clt{The DCC concept arises when one realizes that if the power spectrum underlying wave modes is uniform, the inherent  complexity is lower, whereas if the power spectrum is distributed non-uniformly among the wave modes, the complexity becomes higher \cite{Gleiser:2018jpd}.}
The underlying DCE will be shown to refine the existing analysis of 3-sine-Gordon deformed models, providing new important physical analyses, focusing on two approaches for the deformation procedure of the original sine-Gordon model. The DCE will be also shown to drive the best choice of the parameters involved in the deformation procedure, associated with the DCE global critical points.  
This work is organized as follows: Sect. \ref{sec2} is devoted to introducing the necessary
background for the deformation procedure and constructing the family of deformed sine-Gordon-type models. Sect. \ref{sec3} scrutinizes the role played by the DCE in this context, addressing new subtle aspects. Also, it provides and computes the DCE for the 3-sine Gordon model, this time without approximations but in the context of a deformation procedure without a power expansion. Sect. \ref{sec5} presents the conclusions and perspectives about the important results involving the DCE associated with these models.  

\section{Deformed sine-Gordon-type models}
\label{sec2}
For a scalar field $\upchi=\upchi(t,x)$, one can consider models implemented by the Lagrangian  
\begin{equation}
\mathcal{L}=\frac12 \partial_\mu \upchi \partial^\mu \upchi - V(\upchi),\label{lag1}
\end{equation}
where $V(\upchi)$ denotes the potential governing a scalar field theoretical model.  
Topological solutions can be particularly introduced when static fields are taken into account. In this case, the equation of motion for 
$\upchi=\upchi(x)$ reads  
\begin{equation}
\frac{d^2\upchi}{dx^2}=\frac{dV}{d\upchi}.
\end{equation}
\clt{One searches for configurations of the scalar field that sets forth a minimum $\upchi_{\scalebox{.5}{\textsc{min}}}$ of $V(\upchi$), with boundary conditions  $\lim_{x\to\infty}d\upchi(x)/dx =0$ and $\lim_{x\to-\infty}\upchi(x) = \upchi_{\scalebox{.5}{\textsc{min}}}$.}
Topological solutions usually appear when the potential is non-negative,
which can be written, using the superpotential $W=W(\upchi)$, in the following form, 
\begin{equation}
V(\upchi)=\frac12 W_\upchi^2,\label{sup}
\end{equation}
where $W_\upchi$ stands for\footnote{Hereon the notation $h_\zeta\equiv dh/d\zeta$ will be employed, for any quantity $h$ and any scalar field $\zeta$.} $dW/d\upchi$. It yields the equation of motion \cite{Bazeia:2002xg}
\begin{equation}
\frac{d^2 \upchi}{dx^2}=W_\upchi W_{\upchi \upchi},
\end{equation}
which can be solved when solutions of the first-order ODE  
\begin{equation} \label{firstorder}
\frac{d\upchi}{dx}=W_\upchi
\end{equation}
are employed. 
As Eq. (\ref{sup}) shows that the scalar potential is invariant under $W\mapsto -W$, Eq. \eqref{firstorder}
can be also analyzed in the context of this $\mathbb{Z}_2$-symmetry. Therefore there is a well-defined mapping between the second-order equation of motion into two first-order equations, which in general holds for the topological solution. 
A topological solution of Eq. \eqref{firstorder} on the
topological section ($jk$), has energy minimized to the value $%
E^{jk}_{\scalebox{.5}{\textsc{BPS}}}=|W(\mathring\upchi_j)-W(\mathring\upchi_k)|$, which does represent a saturated inequality depending on the homotopy class of the solution at infinity \cite{Bazeia:2002xg,Epl1}. Here $\mathring \upchi_j$ and $\mathring
\upchi_k$ are \clt{two minima. If these minima are adjacent, clearly $E^{i}_{\scalebox{.5}{\textsc{BPS}}}=|W(\mathring\upchi_{i})-W(\mathring\upchi_{i+1})|$.} It represents the
Bogomol'nyi bound, and the solution of \eqref{firstorder} is named a BPS (Bogomol'nyi--Prasad--Sommerfield)
linearly stable state. By insisting that the BPS bound is saturated, one can come up with a simpler set of differential equations to solve, called the Bogomol'nyi equations. 
Since the superpotential $W(\upchi)$ is $\upchi$-dependent, it
can be used to define the topological behavior from the
topological current density $j_{\scalebox{.6}{\textsc{top}}}^\mu = \varepsilon^{\mu\nu} \partial_\nu W(\upchi).$ 
This definition differs from the standard form of the \clt{topological} current density given by $j^\mu = \varepsilon^{\mu\nu} \partial_\nu \upchi$, \clt{which is also trivially conserved, due to the antisymmetry of the
Levi--Civita tensor. 
One can use $W(\upchi)$ to define the topological current density, yielding the conserved topological charge identical to the energy \cite{Bazeia:1999jt}. This technique of utilizing the superpotential  $W(\upchi)$ can be in some cases more suitable than considering the scalar field $\upchi$ itself, in the construction of the topological current density \cite{Bazeia:1999jt}. The corresponding topological charge is given by
\beq
Q_{\scalebox{.6}{\textsc{top}}} = \int_{\mathbb{R}} \left(\lim_{x\to+\infty}W(\upchi(x))-\lim_{x\to-\infty}W(\upchi(x))\right),\eeq which equals the energy of
the static scalar field configurations. The vacuum sector, which is
identified by time and space independent field configurations, has zero topological charge. The concept of conserved topological
charge, therefore, introduces topological sectors. Distinct  non-vanishing topological charges define different
topological sectors, yielding a quite general classification
scheme for the topological solutions \cite{Bazeia:1999jt,Bazeia:1996cgk}. Besides,  the temporal component $j^0_{\scalebox{.6}{\textsc{top}}} =dW(\upchi(x))/dx$, so that the associated topological charge is directly related to the energy $E=\Delta W=\lim_{x\to+\infty}W(\upchi(x))-\lim_{x\to-\infty}W(\upchi(x))$.} \clt{For more details on useful features of the topological current density, see, e. g., Ref. \cite{Vyas:2014dua}.}

We will focus our attention on the deformation procedure proposed in Refs. \cite{Bazeia:2002xg,Epl1,epjc}. A significant advantage of using deformation approaches consists of straightforwardly generating new models and constructing their topological defects
analytically, from a given model that hereon will be considered the sine-Gordon one. The scalar field model (\ref{lag1}) is supposed to support
topological defects. Deformations of the
sine-Gordon model, whose potential reads 
\begin{equation}
V(\upchi )=\frac{1}{2}\cos ^{2}(\upchi ), \label{cosss}
\end{equation}
 can be therefore implemented. 
This potential is obtained from the superpotential $W(\upchi )=\sin (\upchi)$ and
has an infinite minima, $\mathring{\upchi}=\pm n\pi /2$, for $n\in 2\mathbb{N}+1$, which can be connected by the topological solutions 
\begin{equation}
\upchi(x)=\pm \uptheta (x)\pm k\pi, \label{chi_x}
\end{equation}%
where $k\in\mathbb{Z}$ and $
{\rm gd}(x)\equiv \uptheta (x)=\arcsin [\tanh (x)]$ denotes the
Gudermannian function, whose associated energy reads $E=2$, with dimensionless units being used. \clt{It is worth mentioning that  the  energy
associated to $\upchi(x)$ can be minimized to
\beq
E_{\scalebox{.5}{\textsc{BPS}}}^\pm = \pm \int_{\mathbb{R}} W_\upchi(\upchi)\frac{d\upchi(x)}{dx}\,dx\nonumber
\eeq
when the field configurations obey
\beq
\frac{d\upchi_\pm}{dx} = \pm W_\upchi(\upchi_\pm),\nonumber
\eeq
characterizing BPS states. Hence, the fact that the BPS energy of the minima equals zero comes from the fact that singular points of $W(\upchi)$ are absolute minima of
the potential satisfying $W_\upchi(\upchi)=0$.}

The deformation procedure emulates Eq. (\ref{lag1}), stating that a family of models can be generated by considering a scalar field Lagrangian density defined by 
\begin{equation}
\mathcal{L}=\frac12 \partial_\mu \upphi \partial^\mu \upphi - U(\upphi),
\end{equation}
where $U(\upphi)$ is the potential, governing the new scalar field $\upphi$, written in terms of $V(\upchi)$ as 
\begin{equation}
U(\upphi)=\lim_{\upchi \to f(\upphi)}\frac{V(\upchi)}{f^{2}_\upphi(\upphi)} \,\clt{=\frac12\left(\frac{W'[f(\upphi)]}{f'(\upphi)}\right)^2},\label{start}
\end{equation}
where $f(\upphi)$ is the deformation function. As $\upchi(x)$ is a
static solution of the original sine-Gordon model, then it can be allocated as a solution
of the deformed model whenever the mapping \beq\upphi(x)=f^{-1}[\upchi(x)]\eeq holds. 
\clt{The first-order
equations of the deformed model read
\beq
\frac{d\upphi}{dx}=\pm \mathfrak{W}_\upphi(\upphi) = \pm \frac{W'(f(\upphi))}{f'(\upphi)}.
\eeq}
One can express 
\beq
f(\upphi)=\upphi+\upepsilon g(\upphi),\label{def1}
\eeq where $\upepsilon$ is a small
parameter that allows a power expansion with respect to it \cite{Bazeia:2002xg,Epl1}, and $g(\upphi)$ is, a priori, an arbitrary function regulating the deformation, being compatible with the $\upepsilon$-expansion. Although the deformation procedure holds for a general potential, models controlled by the superpotential $W(\upchi)$ will be approached, where $V(\upchi)=\frac12 W_\upchi^2.$ This expression can be emulated for the new potential as  
\begin{equation} \label{calWp}
U(\upphi)=\frac12 \mathfrak{W}_\upphi^2,
\end{equation}
where defining the new superpotential \cite{Bazeia:2002xg} \clt{as an $\upepsilon$-deformation of the superpotential $W(\upchi)=W(f(\upphi))$}, as 
\beq
\mathfrak{W}=\mathfrak{W}(\upphi)=W(\upchi)+\upepsilon
W_\upepsilon(\upphi),\label{s1}
\eeq yields  
\begin{equation}
\mathfrak{W}_\upphi = W_\upphi - \upepsilon \left( W_\upphi g_\upphi - g W_{\upphi\upphi}\label{s2}
\right).
\end{equation}
Comparing Eqs. (\ref{s1}) and (\ref{s2}) makes it possible to write 
\begin{equation}\label{ss3}
W_\upepsilon(\upphi)=\int^\upphi \left(gW_{\phi\phi}-W_\phi g_\phi\right)\,d\phi\,,
\end{equation}
which can be utilized to derive the minima of the potential by analyzing the deformed ODE 
\begin{equation}
W_\upphi=\upepsilon \left( W_\upphi g_\upphi - gW_{\upphi\upphi} \right).
\end{equation}
The minima can be used to find the corresponding energy in each topological
sector, in the form \clt{$E_{i}=|\mathfrak{W}_i-\mathfrak{W}_{i+1}|$, where $\mathfrak{W}_i\equiv \mathfrak{W}(\upphi_i)$, given $\upphi_{i+1}$ and $\upphi_i$ representing two
adjacent minima of the $U(\upphi)$}. Also, taking into account Eq. (\ref{def1}), up to the first order in $%
\upepsilon$, the solution of the new model can be expressed as \cite{epjc}  
\beq
\upphi(x)=\upchi(x)-\upepsilon g(\upchi(x)).
\eeq

\begin{figure}[H]\bec
\includegraphics[scale=0.35]{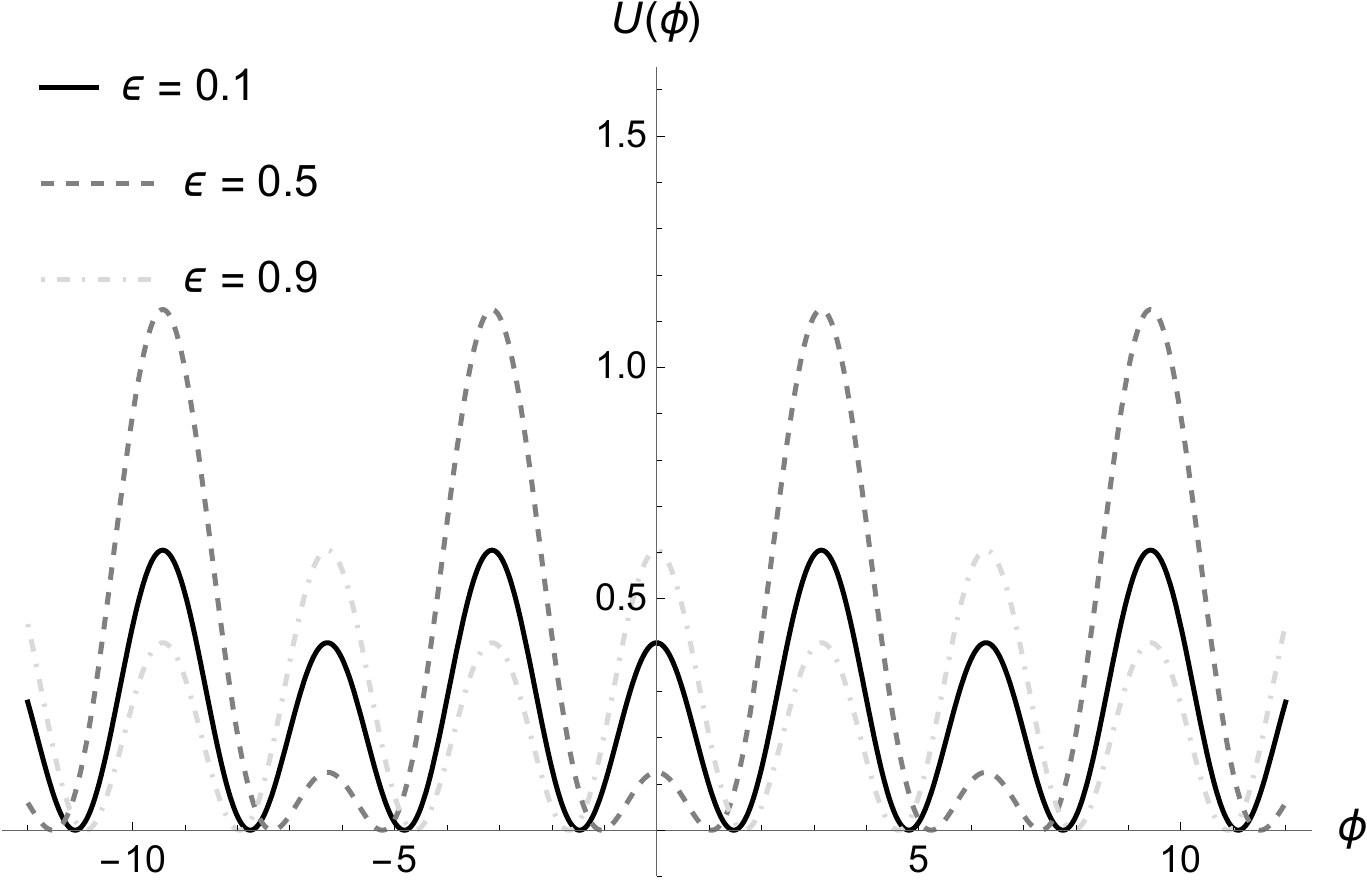}\caption{Profile of the deformed potential $U(\upphi)$, for different values of $\upepsilon$, for $a=1$.}
\label{pott1}\eec
\end{figure}

\begin{figure}[H]\bec
\includegraphics[scale=0.35]{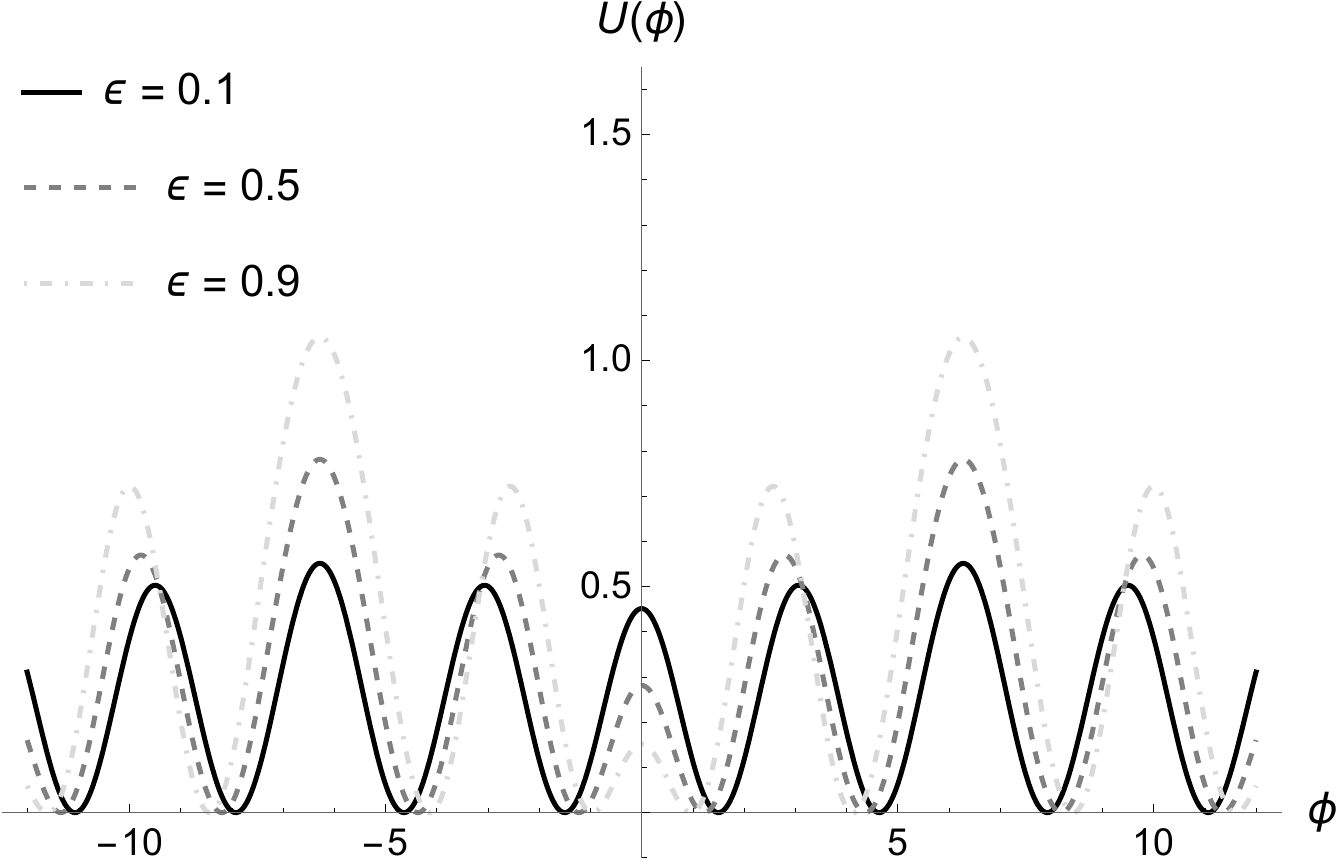}\caption{Profile of the deformed potential $U(\upphi)$, for different values of $\upepsilon$, for $a=2$.}\label{pott2}\eec
\end{figure}

To introduce a new family of sine-Gordon-type models, Refs. \cite{Bazeia:2002xg,Epl1,epjc} introduced the original sine-Gordon given by Eq. \eqref{cosss} and implemented the deformation
function as 
\begin{equation}
f_{a}(\upphi)=\upphi +\upepsilon g_{a}(\upphi),
\label{func_deform}
\end{equation}
where, for deforming the sine-Gordon model, one defines \cite{epjc}
\beq
g_{a}(\upphi)\equiv \sin \left( \frac{\upphi }{a}\right).\eeq
The deformation
function \eqref{func_deform} is parametrized by two real parameters, $a\neq0$ and $\upepsilon$, with small $\upepsilon$. A straightforward advantage of this protocol is the immediate implementation yielding potentials that have interest in relevant applications. Besides, the small parameter $\upepsilon$ identifies topological sector solutions of the first-order equation, with respective energy densities associated with the deformed models. The deformation
function defines a family of sine-Gordon-type models described by the potential \eqref{calWp},
where the superpotential $\mathfrak{W}$ has derivative with respect to the scalar field $\upphi$ given by  
\begin{equation}
\mathfrak{W}_{\upphi }\!=\!\cos (\upphi )-\!\upepsilon \!\left[ \sin (\upphi )\sin
\!\left(a^{-1}{\upphi }\right) \!+a^{-1}\!\cos (\upphi )\cos \!\left( 
a^{-1}{\upphi }\right) \!\right] . \label{wsg}
\end{equation}%
An additional term arises, being responsible to shift the minima of the potential to new ones, given by 
\begin{equation}
\upphi _{\pm }^{\scalebox{.65}{\textsc{min}}}=\pm \frac{k\pi }{2}\mp\upepsilon \sin \left( \frac{%
n\pi }{2a}\right),\,\,\,\,\,k\in2\mathbb{N}+1, \label{minima}
\end{equation}
also modifying the maxima as
\begin{equation}
\upphi _{\pm }^{\scalebox{.65}{\textsc{max}}}=\pm p\pi \pm\frac{\upepsilon}{a^{2}} (a^{2}-1)\sin
\left( \frac{p\pi }{a}\right),\,\,\,\,\,p\in\mathbb{N}.
\label{maxima}
\end{equation}
When $a\in\mathbb{Z}$, the number of distinct topological sectors, 
labeled by $p$, equals $a+1$. 
Therefore the deformed potential is realized to have periodicity $2\pi a$. Taking into account $a=1$ yields the 2-sine-Gordon, whereas, for $a=2$, the 3-sine-Gordon model can be obtained, with immediate generalizations for each value of $a$. For $a\neq
1 $, the new superpotential reads  
\begin{eqnarray}
\mathfrak{W}=\sin (\upphi )\! &-&\!\frac{\upepsilon }{2}\left\{ \frac{(a+1)}{a-1%
}\sin \left[ \frac{(a-1)}{a}\upphi \right] +\frac{(1-a)}{a+1}\sin \left[ \frac{(a+1)}{a}\upphi \right]
\right\}.
\end{eqnarray}
For the particular case where $a=1$, the new superpotential has the straightforward expression
\begin{equation}
\mathfrak{W}(\upphi)=\sin (\upphi)-\upepsilon \upphi.\label{new1}
\end{equation}
\clt{Based on the very definition of the deformed superpotential in Eq. \eqref{calWp}, the corresponding deformed potential is depicted in Figs. \ref{pott1} and \ref{pott2}, for different values of the parameter $a$.}

One can then employ $\mathfrak{W}$ and the minima \eqref{minima} to
compute the energy associated with a topological defect corresponding to each topological sector. The general expression for $%
\mathfrak{W}$ in each minimum is \cite{epjc}
\begin{equation}
\mathfrak{W}(\upphi _{\pm }^{\scalebox{.65}{\textsc{min}}})=\pm (-1)^{{\scalebox{0.75}{\textsc{$(n-1)/2$}}}}\left[ 1-\frac{2a}{%
a^{2}-1}\cos \left( \frac{n\pi }{2a}\right) \upepsilon \right] ,
\label{Wsuper}
\end{equation}
which yields the particular case regarding the standard sine-Gordon model for $a=1$,  
\begin{equation}
\mathfrak{W}(\upphi _{\pm }^{\scalebox{.65}{\textsc{min}}})=\pm (-1)^{{\scalebox{0.75}{\textsc{$(n-1)/2$}}}}\mp\upepsilon 
\frac{n\pi }{2}.
\end{equation}
Also, topological solutions associated with the first-order equation 
\eqref{firstorder} can be derived using the inverse deformation
function. Eq. \eqref{func_deform} is hence employed to yield \cite{epjc} 
\begin{equation}
\upphi (x)=\upphi _{0}(x)-\upepsilon \sin \left[ \frac{1}{a}\upphi _{0}(x)\right],
\end{equation}%
where $\upphi _{0}(x)=\upchi (x)$ denotes the solutions \eqref{chi_x} of the original sine-Gordon model. The associated energy density reads  
\begin{equation}
\rho (x)=\rho _{0}(x)\left[ 1-\frac{2\upepsilon }{a}\cos \left( \frac{\upphi
_{0}(x)}{a}\right) \right] ,
\end{equation}%
with $\rho _{0}(x)=\text{sech}^{2}(x)$ representing the standard energy density associated with sine-Gordon classical solutions.

Now the case of the 3-sine-Gordon model can be considered, which is
obtained with $a=2$. There are three kinds of topological
sectors. The sectors of the first kind are described by the following
solutions \cite{epjc}
\begin{equation}
\upphi _{1}(x)=\uptheta (x)+4\ell\pi -\frac{\sqrt{2}}{2} {\rm sgn}(x) 
\left[ 1-\text{sech}(x)\right]^{1/2}\upepsilon, \label{eq1}
\end{equation}%
where $\ell$ is an integer, that
connect the minima 
\bes
\clt{
\beq
M_1 &=&- \frac{\pi }{2}+4\ell\pi + \frac{\sqrt{2}}{2}\upepsilon\,,\label{mm1}\\
\mathring{M}_1 &=&+ \frac{\pi }{2}+4\ell\pi \pm \frac{\sqrt{2}}{2}\upepsilon\,.\label{mm2}
\eeq}
\ees
\clt{The first topological sector (\ref{eq1}) and the respective minima 
(\ref{mm1}, \ref{mm2}) are depicted in Fig. \ref{min1}.
\begin{figure}[H]
\bec
\includegraphics[scale=0.41]{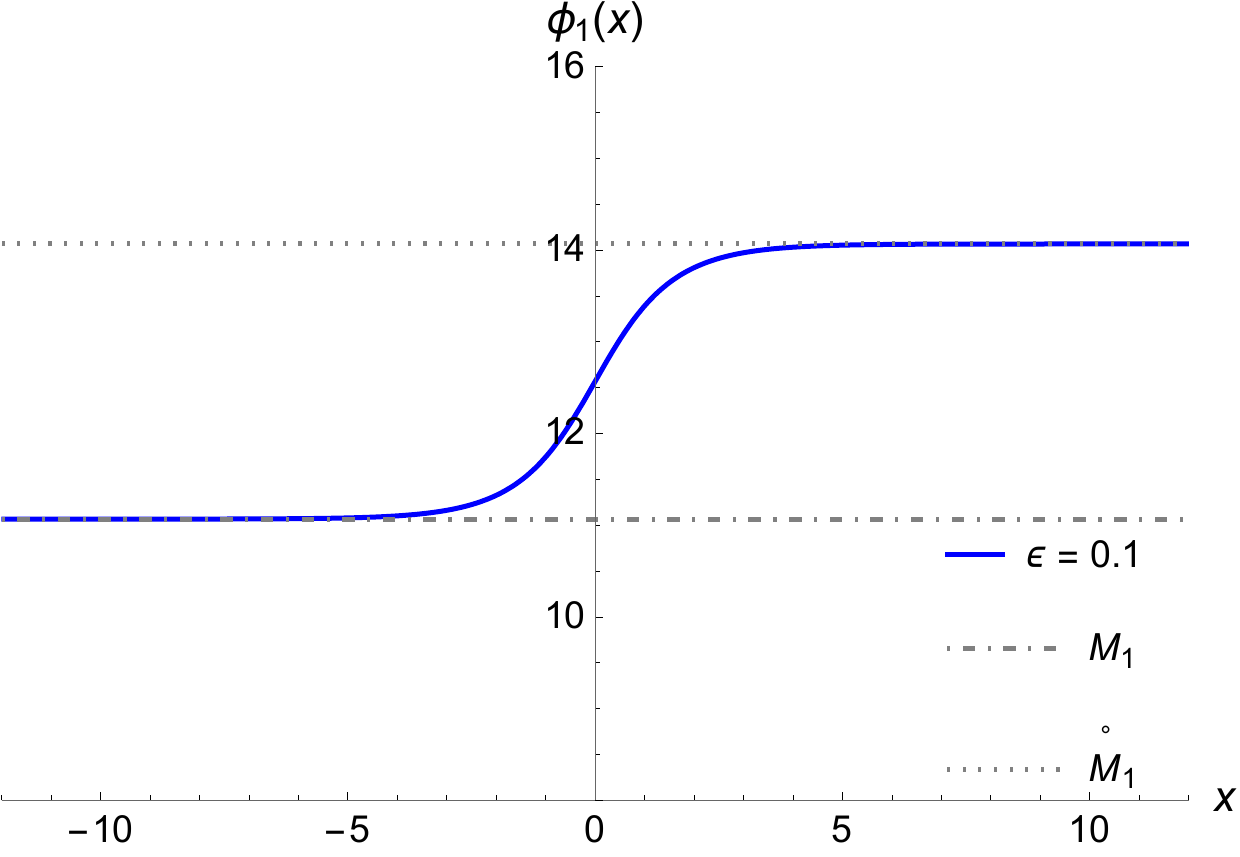}\caption{Profile of the topological sector $\protect\upphi_1(x)$, for  
$\protect\upepsilon=0.1$ and $\ell=1$.}
\label{min1}\eec
\end{figure}}
The sectors of the second kind are described by the solutions 
\begin{equation}
\upphi_{2}(x)=\uptheta (x)+(4\ell+1)\pi -\frac{\sqrt{2}}{2}
\left[ 1-\text{sech}(x)\right]^{1/2}\upepsilon, \label{eq2}
\end{equation}%
which connect the minima 
\bes
\clt{\beq
M_2=\frac{\pi }{2}+4(\ell+1)\pi -\frac{\sqrt{2}}{2}\upepsilon,\label{m22}\\
\mathring{M}_2=-\frac{\pi }{2}+4(\ell+1)\pi -\frac{\sqrt{2}}{2}\upepsilon.\label{m2}
\eeq}
\ees
\clt{The second topological sector (\ref{eq2}) and the respective minima 
(\ref{m22}, \ref{m2}) are depicted in Fig. \ref{min2}.
\begin{figure}[H]
\bec\includegraphics[scale=0.41]{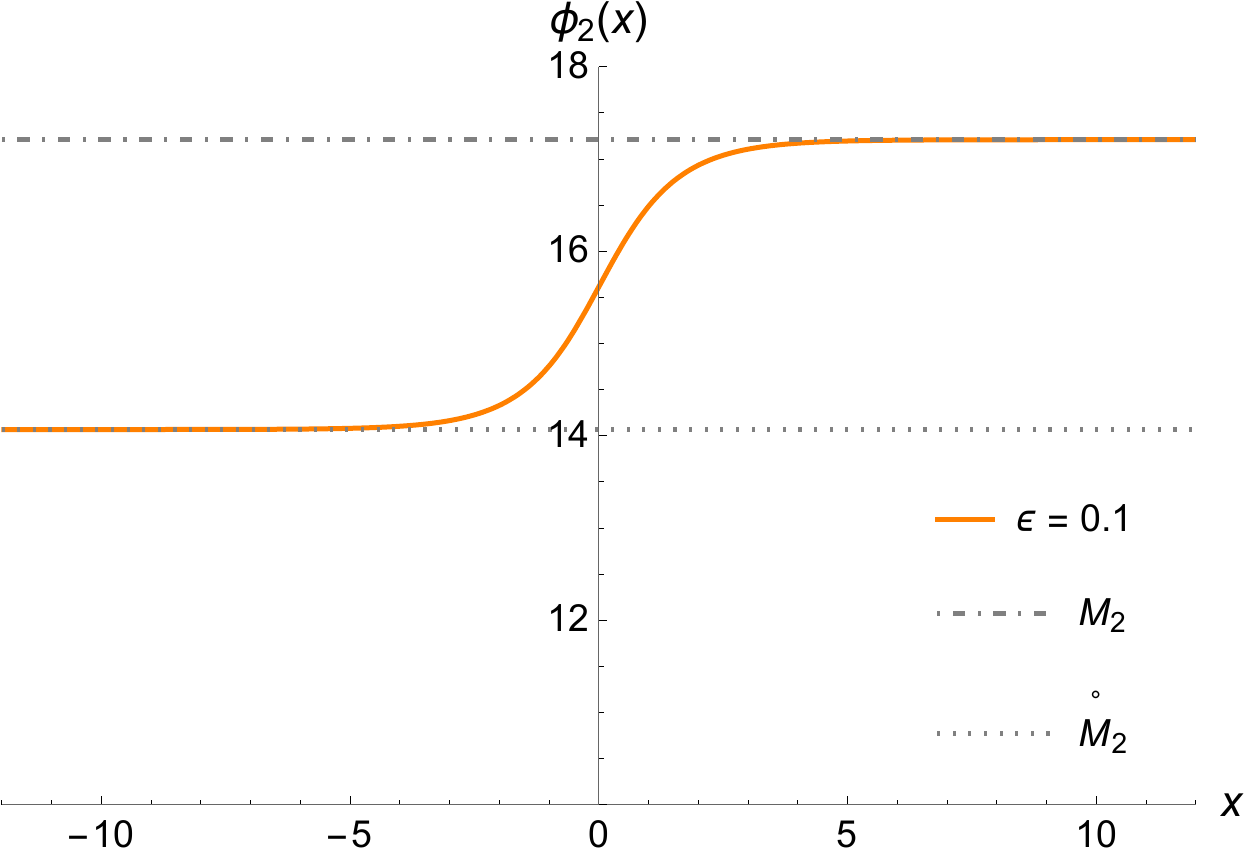}\caption{Profile of the topological sector $\protect\upphi_2(x)$, for  
$\protect\upepsilon=0.1$ and $\ell=1$.}
\label{min2}\eec
\end{figure}}
The third kind of topological sector is described by the solutions 
\begin{equation}
\upphi _{3}(x)=\uptheta (x)+2\ell\pi +\frac{\sqrt{2}}{2}{\rm sgn}(x)
\left[ 1-\text{sech}(x)\right]^{1/2}\upepsilon, \label{eq3}
\end{equation}%
which connect the minima 
\bes
\beq
M_3&=&-\frac{3\pi }{2}+4\ell\pi +\frac{\sqrt{2}}{2}\upepsilon,\label{m3}\\
\mathring{M}_3&=&-\frac{5\pi }{2}+4\ell\pi +\frac{\sqrt{2}}{2}\upepsilon,\label{m33}
\eeq
\ees 
\clt{The third topological sector (\ref{eq3}) and the respective minima 
(\ref{m3}, \ref{m33}) are displayed in Fig. \ref{min3}.
\begin{figure}[H]
\bec
\includegraphics[scale=0.41]{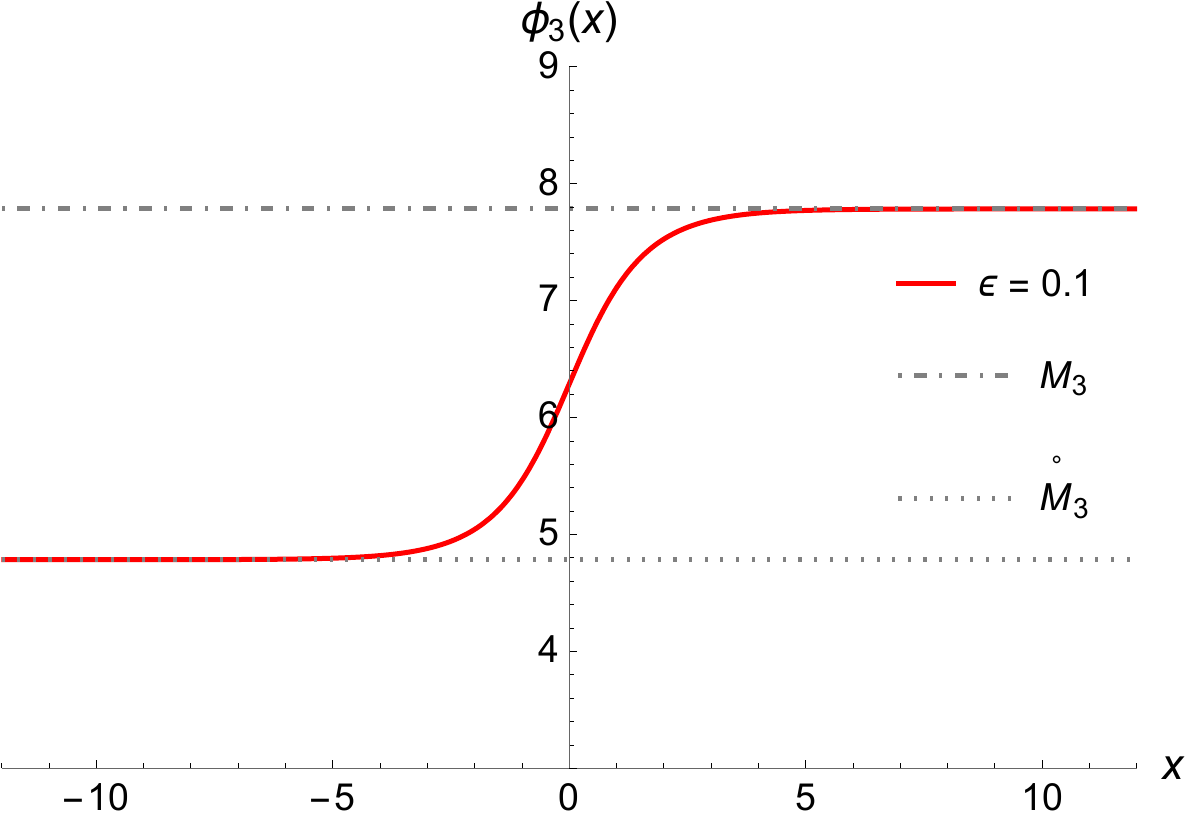}\caption{Profile of the topological sector $\protect\upphi_3(x)$, for  
$\protect\upepsilon=0.1$ and $\ell=1$.}
\label{min3}\eec
\end{figure}}

The energy densities of the solutions corresponding to the three distinct
topological sectors are given by \cite{epjc}
\begin{subequations}
\begin{eqnarray}
\hspace*{-0.4cm}\!\!\!\!\rho _{1}(x)\!&\!=\!&\!\left( 1\!-\frac{\sqrt{2}}{2}
\left[ 1+\text{sech}(x)\right]^{1/2}\upepsilon\right)\text{sech}^{2}(x), \label{ro1} \\
\hspace*{-0.4cm}\rho _{2}(x)\!&\!=\!&\!\left( 1\!+\!\frac{\sqrt{2}}{2}{\rm sgn}(x)
\left[ 1-\text{sech}(x)\right]^{1/2}\upepsilon\right)\text{sech}^{2}(x),\label{ro2}\\
\hspace*{-0.4cm}\rho _{3}(x)\!&\!=\!&\!\left( 1\!+\frac{\sqrt{2}}{2}
\left[ 1+\text{sech}(x)\right]^{1/2}\upepsilon
\right)\text{sech}^{2}(x),\label{rho123}
\end{eqnarray}%
\end{subequations}
\clt{whose plots, for various values of $\upepsilon$, are described in Figs. \ref{eed1} -- \ref{eed3}. 
\begin{figure}[H]\bec
\includegraphics[scale=0.4]{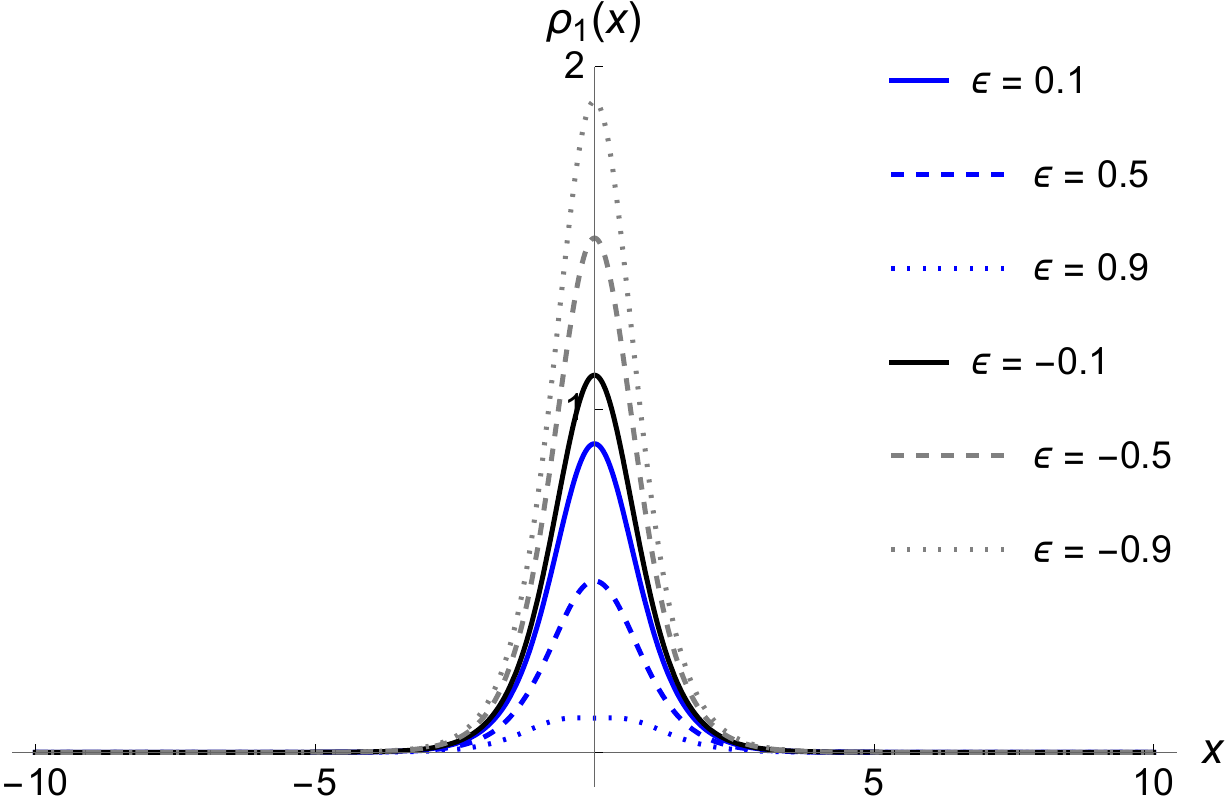}\caption{Plots of the energy density $\rho _{1}(x)$ of the first topological sector, for different values of $\upepsilon$.}
\label{eed1}\eec
\end{figure}
\begin{figure}[H]\bec
\includegraphics[scale=0.4]{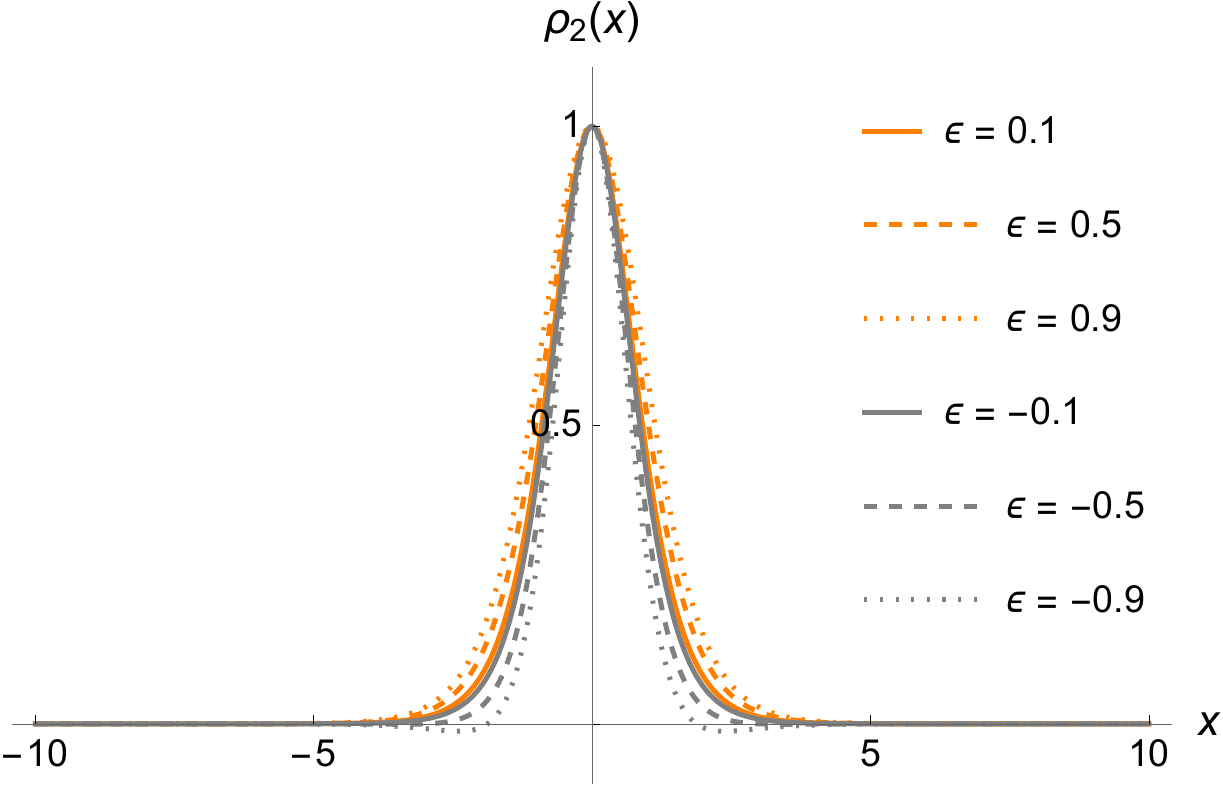}\caption{Plots of the energy density $\rho _{2}(x)$ of the second topological sector, for different values of $\upepsilon$.}
\label{eed2}\eec
\end{figure}
\begin{figure}[H]\bec
\includegraphics[scale=0.4]{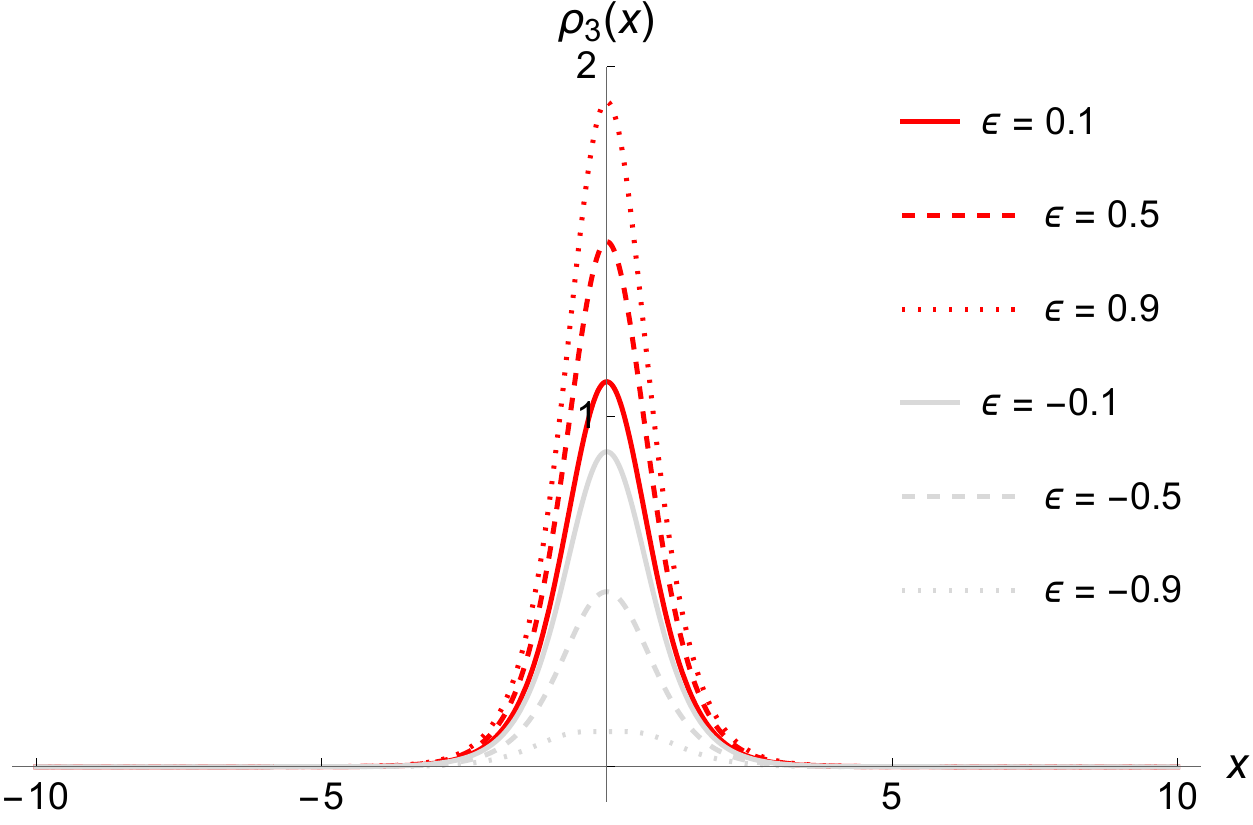}\caption{Plots of the energy density $\rho _{3}(x)$ of the third topological sector, for different values of $\upepsilon$.}
\label{eed3}\eec
\end{figure}}
\clt{One can realize the localization of the energy densities around $x=0$.  Fig. \ref{eed2} illustrates the energy density $\rho _{2}(x)$ of the second topological sector presenting a small variation with respect to the deformation parameter $\upepsilon$. Conversely, the energy densities $\rho _{1}(x)$ and $\rho _{3}(x)$, respectively in Figs. \ref{eed1} and \ref{eed3} reflect a  vast  variation with respect to $\upepsilon$. }

Under the $\mathbb{Z}_2$-symmetry $\upepsilon \mapsto -\upepsilon$, the first and third topological sectors are interchanged and the energy densities related to the first and third sectors are mapped into each other as well. 
Eq. \eqref{Wsuper} yields the corresponding energies, respectively corresponding to each topological sector (\ref{ro1}) -- (\ref{rho123}), 
\begin{equation}
E_{1}=2-\frac{4\sqrt{2}}{3}\,\upepsilon,\qquad\,\,\,\,E_{2}=2,\qquad \,\,\,\,E_{3}=2+\frac{%
4\sqrt{2}}{3}\,\upepsilon . \label{e123}
\end{equation}
{{To ensure energy positivity, in the first and third topological sectors it must be imposed that $E_{1}>0$ and $E_{3}>0$. Hence it implies the bound on $\upepsilon$ for these two sectors
\[
-\frac{3\sqrt{2}}{4}<\upepsilon<\frac{3\sqrt{2}}{4}.
\]
It is worth emphasizing that the second topological sector presents energy $E_{2}=2$. Thus there is no restriction on the 
parameter $\upepsilon$ and the energy is degenerate. 

Figs. \ref{fig1} -- \ref{fig3} illustrate the profile of the respective solution for each
topological sector. The variation of the configurations according to $%
\upepsilon$ can be realized likewise. 
\begin{figure}[H]\bec
\includegraphics[scale=0.41]{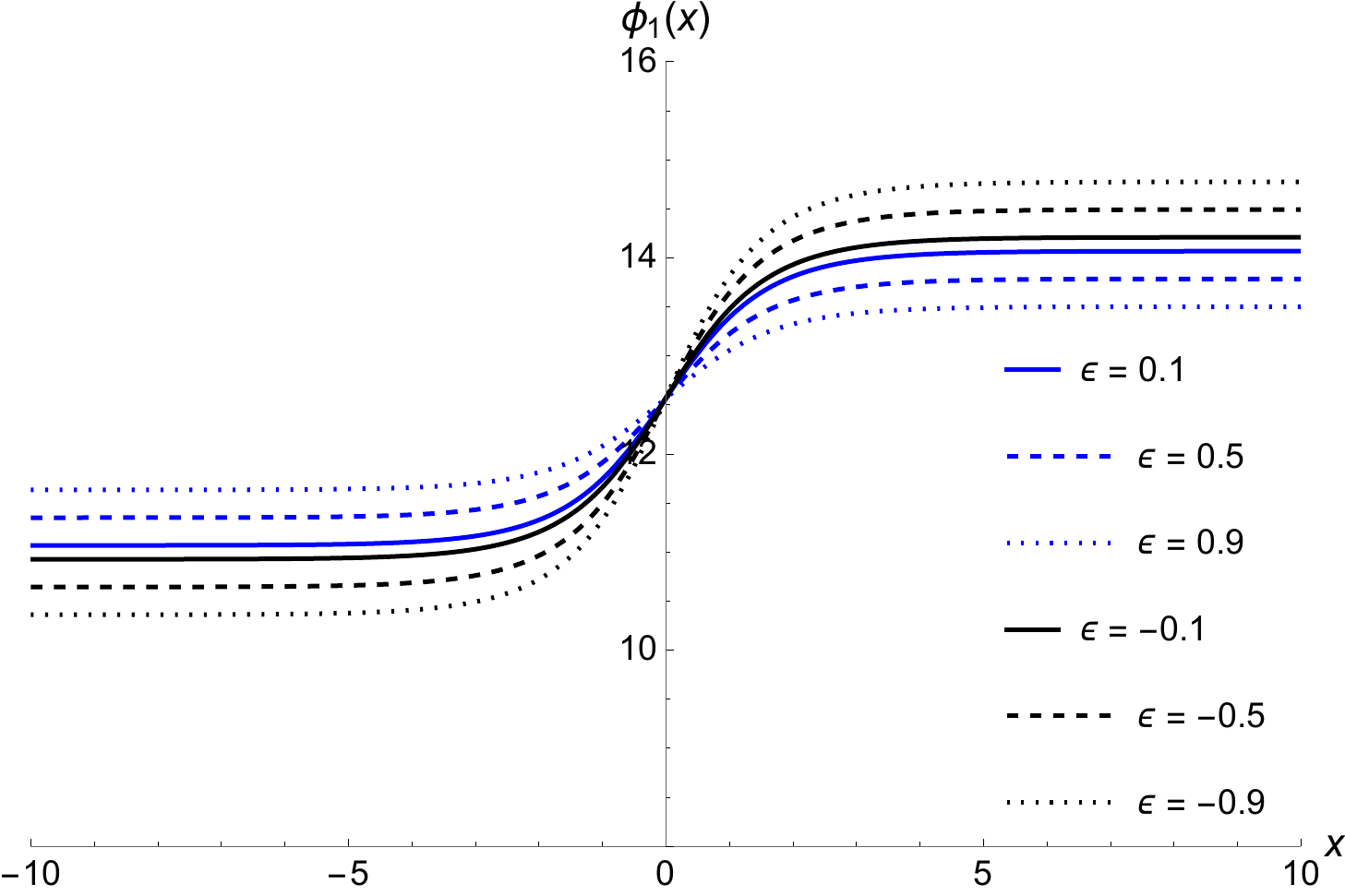} \caption{Profile of the topological sector $\protect\upphi_1(x)$, for  
$\protect\upepsilon=0.1$, $\protect\upepsilon=0.5$, and $\protect\upepsilon=0.9$ (blue) and $\protect\upepsilon=-0.1$, $\protect\upepsilon=-0.5$, and $\protect\upepsilon=-0.9$ (black).}\label{fig1}\eec
\end{figure}
One can realize from Fig. \ref{fig1} that the scalar field (\ref{eq1}), regarding the first topological sector, has lower [higher] absolute asymptotic values, for $x> 0$ [$x<0$], when the positive values of $\upepsilon$ are respectively compared to the negative values of $\upepsilon$. A similar behavior holds for the scalar fields regarding the second and third topological sectors, respectively in Eqs. (\ref{eq2}, \ref{eq3}), illustrated in Figs. \ref{fig2} and \ref{fig3}. 
\begin{figure}[H]\bec
\includegraphics[scale=0.41]{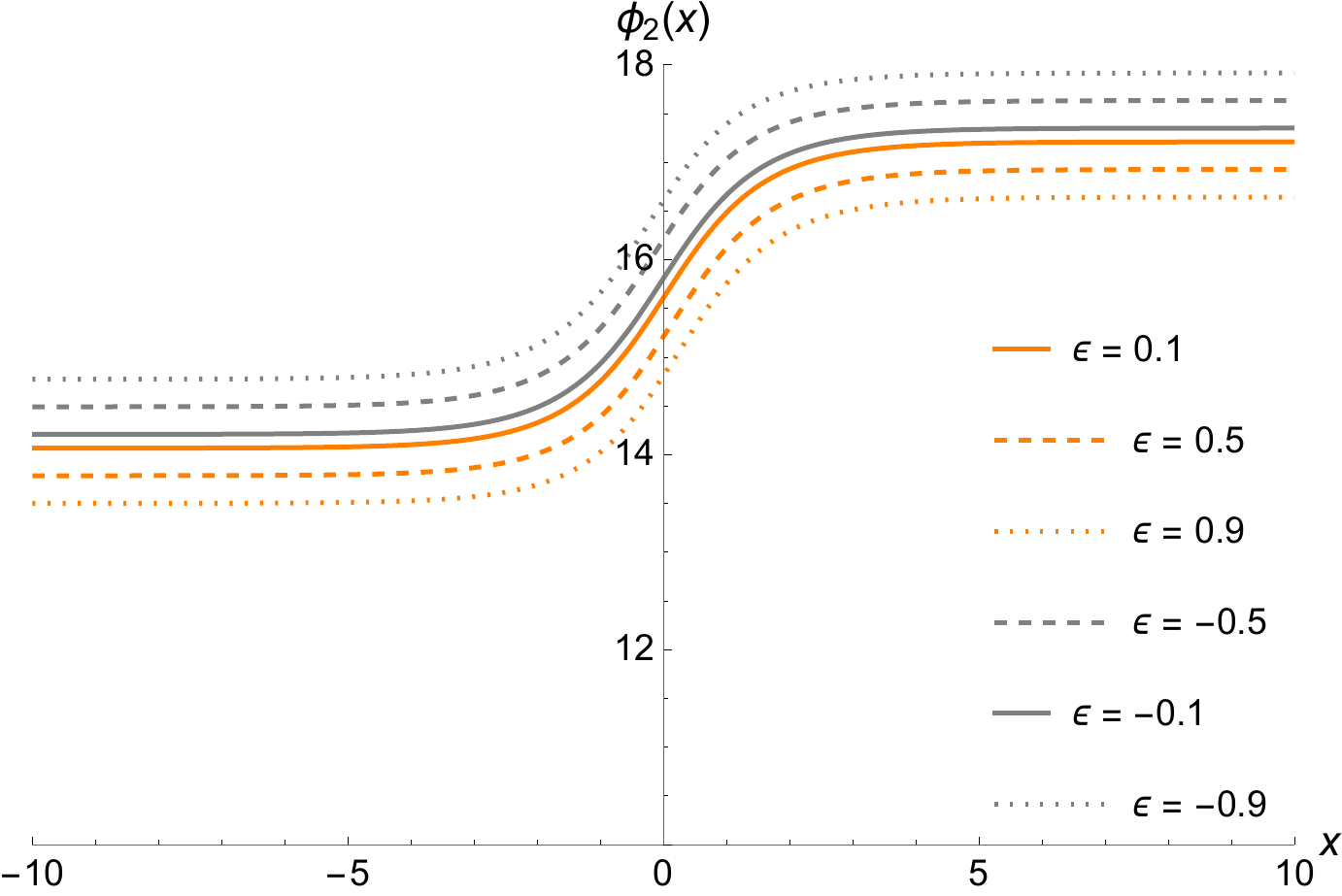}\qquad
\caption{Profile of the topological sector $\protect\upphi_2(x)$, for  
$\protect\upepsilon=0.1$, $\protect\upepsilon=0.5$, and $\protect\upepsilon=0.9$ (orange) and $\protect\upepsilon=-0.1$, $\protect\upepsilon=-0.5$, and $\protect\upepsilon=-0.9$ (dark-grey).}\label{fig2}\eec
\end{figure}

\begin{figure}[H]\bec
\includegraphics[scale=0.41]{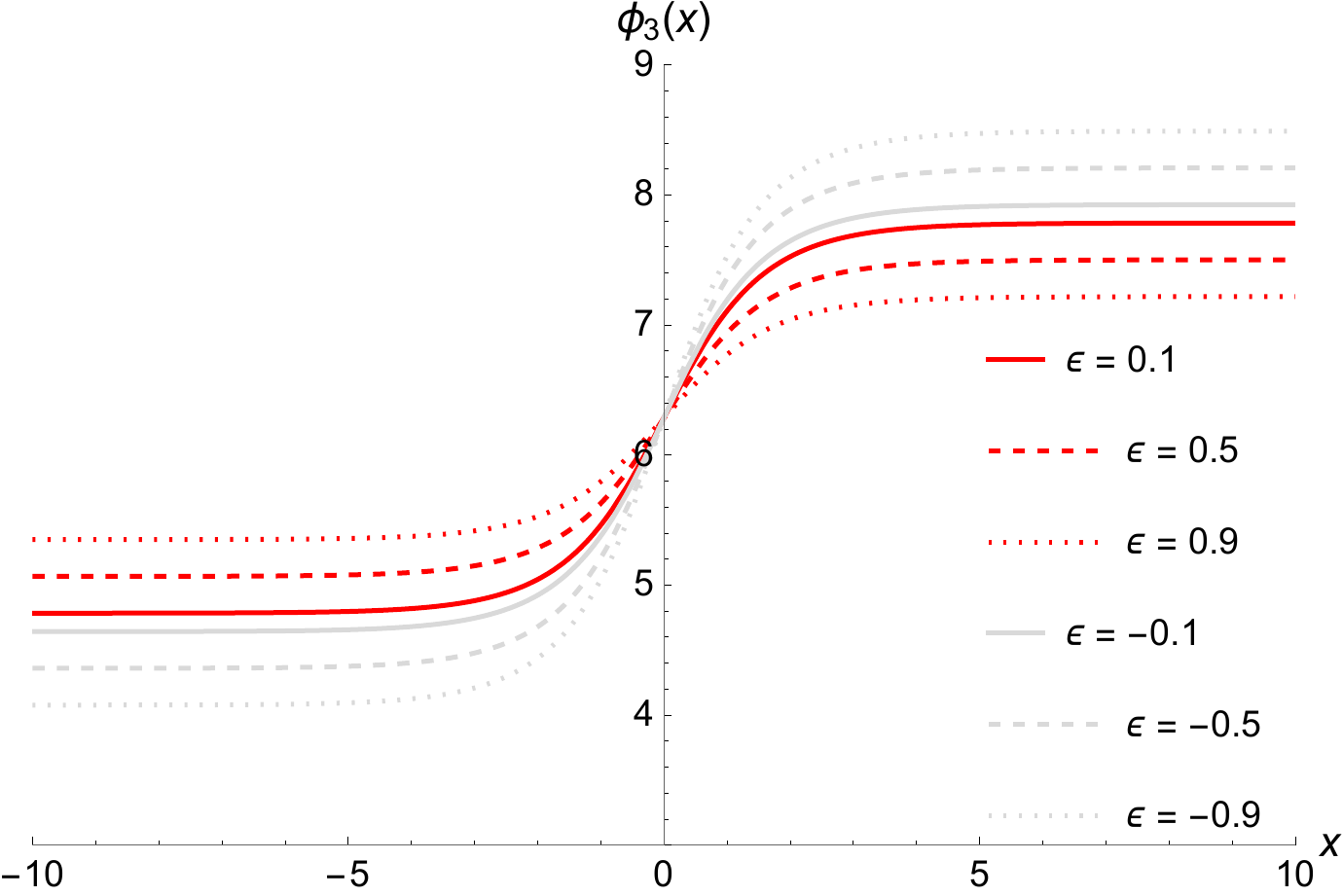}\qquad
\caption{Profile of the topological sector $\protect\upphi_3(x)$, for  
$\protect\upepsilon=0.1$, $\protect\upepsilon=0.5$, and $\protect\upepsilon=0.9$ (red) and $\protect\upepsilon=-0.1$, $\protect\upepsilon=-0.5$, and $\protect\upepsilon=-0.9$ (light-grey).}\label{fig3}\eec
\end{figure}


\section{DCE and the deformed 3-sine-Gordon model}
\label{sec3}
Gleiser and Stamatopoulos  
proposed a precise picture of the DCE 
for discussing the structure of localized solutions in classical field theories \cite{gs12a,gs12b}. The DCE is here addressed, using the deformed sine-Gordon field configurations, where
braneworld scenarios can be studied. Firstly, the DCE setup will be briefly revisited. For it, the Napierian logarithm will be employed, with the DCE quantified in the natural unit of entropy (nat). The DCE regards the number of bits needed to encode localized physical configurations with the best compression.
The protocol for computing the DCE consists of evaluating the Fourier transform of the energy density  
\begin{eqnarray}
{\scalebox{.85}{\textsc{$\rho$}}}(\mathsf{k}) =\frac1{\sqrt{2\pi}} \int_{\mathbb{R}}{\scalebox{.85}{\textsc{$\rho$}}}(\mathsf{r})e^{-i\mathsf{k}\mathsf{r}}d\mathsf{r},\label{ftrans}
\end{eqnarray} Distinct wavemodes are regulated by the modal fraction, whose expression is given by  
\begin{eqnarray}\label{modall}
m_{\scalebox{.8}{\textsc{$\rho$}}}(\mathsf{k}) = \frac{|{\scalebox{.85}{\textsc{$\rho$}}}(\mathsf{k})|^2}{\int_{\mathbb{R}}|{\scalebox{.85}{\textsc{$\rho$}}}({\mathtt{k}})|^2d\mathtt{k}},\label{modalf}
\end{eqnarray} also circumventing the issue involving complex-valued Fourier transforms. 
The DCE is therefore given by 
\cite{gs12b,Gleiser:2018kbq,Gleiser:2018jpd}, 
\begin{eqnarray}
S_{\scalebox{.55}{\textsc{DCE}}}[{\scalebox{.85}{\textsc{$\rho$}}}] = - \int_{\mathbb{R}}\accentset{\star}{m}_{{\scalebox{.81}{\textsc{$\rho$}}}}({{\mathtt{k}}})\log\accentset{\star}{m}_{{\scalebox{.81}{\textsc{$\rho$}}}}({\mathtt{k}})\,d\mathtt{k},\label{ce1}
\end{eqnarray}
where $\accentset{\star}{m}_{\scalebox{.8}{\textsc{$\rho$}}}(\mathsf{k})=m_{\scalebox{.8}{\textsc{$\rho$}}}(\mathsf{k})/m_{\scalebox{.8}{\textsc{$\rho$}}}^{\scalebox{.52}{\textsc{sup}}}(\mathsf{k})$, for $m_{\scalebox{.8}{\textsc{$\rho$}}}^{\scalebox{.52}{\textsc{sup}}}(\mathsf{k})$ denoting the modal fraction supremum. The DCE estimates the dynamical degree of order associated with solutions of the equations of motion regulating the scalar field. 
The configurational stability of the system portrayed by the energy density corresponds to the minima of the DCE. 

\clt{Now one can follow the important contributions in Ref. \cite{Gleiser:2018jpd} and additionally introduce the differential configurational complexity (DCC). For it, one must first specify the modal fraction, normalized by the maximum mode contribution,
\begin{eqnarray}\label{mf2}
\mathsf{m}_{\scalebox{.8}{\textsc{$\rho$}}}(\mathsf{k}) = 
\frac{|{\scalebox{.85}{\textsc{$\rho$}}}(\mathsf{k})|^2}{\textsc{max}_{{\mathtt{k}}}|{\scalebox{.85}{\textsc{$\rho$}}}({\mathtt{k}})|^2}.
\end{eqnarray}
Given a specific power spectrum, the relative contribution 
of distinct wave modes can be quantified by the modal fraction (\ref{mf2}). In fact, this concept arises when one realizes that if the power spectrum underlying wave modes is uniform, the inherent  complexity is lower, whereas if the power spectrum is distributed non-uniformly among the wave modes, the complexity becomes higher.
The normalization with the maximum mode yields the positivity of the DCC, which is therefore given by 
\cite{Gleiser:2018jpd}, 
\begin{eqnarray}\label{dcc}
S_{\scalebox{.55}{\textsc{DCC}}}[{\scalebox{.85}{\textsc{$\rho$}}}] = - \int_{\mathbb{R}}{\mathsf{m}}_{{\scalebox{.81}{\textsc{$\rho$}}}}({{\mathtt{k}}})\log\mathsf{m}_{{\scalebox{.81}{\textsc{$\rho$}}}}({\mathtt{k}})\,d\mathtt{k}.
\end{eqnarray}
As discussed in Ref.  \cite{Gleiser:2018jpd}, the DCC \eqref{dcc} equals zero whenever the wave modes carry the same weight. For example, uncorrelated noise presents a power spectrum of uniform type, yielding a maximal DCE, with DCC vanishing, whereas for a plane wave, meanwhile, both DCE and DCC vanish. It reinforced the influential interpretation posed in Ref.  \cite{Gleiser:2018jpd} of the DCC as a useful measure of shape complexity. }

\subsection{\clt{DCE and DCC of the deformed 3-sine-Gordon model}}
\label{ssec3}
Here we aim to determine the \clt{DCE and the DCC, together with the} subsequent information entropic features of the topological sectors provided by
Eqs. (\ref{eq1}, \ref{eq2}, \ref{eq3}), with respective energy
densities (\ref{ro1}) -- (\ref{rho123}) and their corresponding energies (\ref{e123}). 
 It is worth emphasizing that the second sector $\upphi _{2}(x)$
presents degenerate energy with respect to the parameter $\upepsilon $.
\bigskip 
\begin{figure}[H]\bec
\includegraphics[scale=0.42]{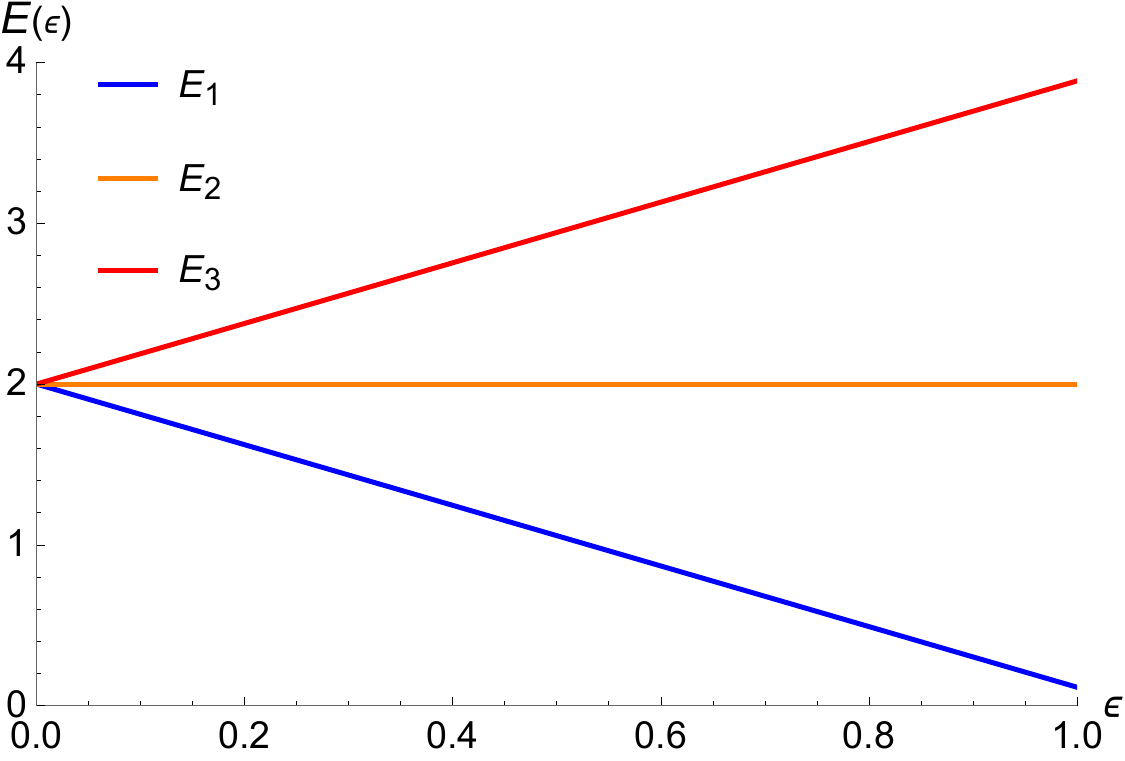}.
\caption{Energy associated to the three distinct topological sectors, (\ref{e123})
with respect to the deformation parameter $\upepsilon $.}\eec
\end{figure}
We can apply to each topological sector the DCE protocol, with corresponding Fourier transform of the energy density (\ref{ftrans}), for computing the modal fraction (\ref{modalf}), and subsequently, the DCE \eqref{ce1}. The analytical results are explicitly expressed in Appendix \ref{aa}.

Now, to completely specify the modal fraction $m_{\scalebox{.8}{\textsc{$\rho_j$}}}(\mathsf{k})$ (\ref{modalf}), for each topological sector $i=1,2,3$, the value $\int_{\mathbb{R}} |{\scalebox{0.95}{\textsc{$\rho_j$}}}(\mathtt{k})|^2\,d\mathtt{k}$ must be computed. Alternatively, one may use the Plancherel
theorem, 
\begin{equation}
\int_{\mathbb{R}} \left\vert {\scalebox{0.95}{\textsc{$\rho_j$}}}(\mathtt{k})\right\vert^{2}\,d\mathtt{k}=\int_{\mathbb{R}} \left\vert
\rho_j(\mathsf{r})\right\vert ^{2}\,d\mathsf{r}. \label{plancherel}
\end{equation}
Hence, the denominator entering the modal fraction expression (\ref{modalf}) reads, respectively for each topological sector, the following analytical functions of the deformation parameter $\upepsilon$: 
\bes
\begin{eqnarray}
\int_{\mathbb{R}} |{\scalebox{0.95}{\textsc{$\rho_1$}}}(\mathtt{k})|^2\,d\mathtt{k} &=& \frac{4}{3}-\frac{64\sqrt{%
2}}{35}\upepsilon + \frac{1}{48}(32+9\pi)\upepsilon^2, \\
\int_{\mathbb{R}} |{\scalebox{0.95}{\textsc{$\rho_2$}}}(\mathtt{k})|^2\,d\mathtt{k} &=& \frac{1}{48}(64 +
32\upepsilon^2-9\pi \upepsilon^2), \\
\int_{\mathbb{R}} |{\scalebox{0.95}{\textsc{$\rho_3$}}}(\mathtt{k})|^2\,d\mathtt{k} &=& \frac{4}{3}+\frac{64\sqrt{%
2}}{35}\upepsilon + \frac{1}{48}(32+9\pi)\upepsilon^2.
\end{eqnarray}
\ees
Figs. \ref{fig8} -- \ref{fig12} illustrate the modal fraction profile for the three topological sectors (\ref{ro1}) -- (\ref{rho123}), for various values of the $\upepsilon$ parameter. 
\begin{figure}[H]
\bec\includegraphics[scale=0.41]{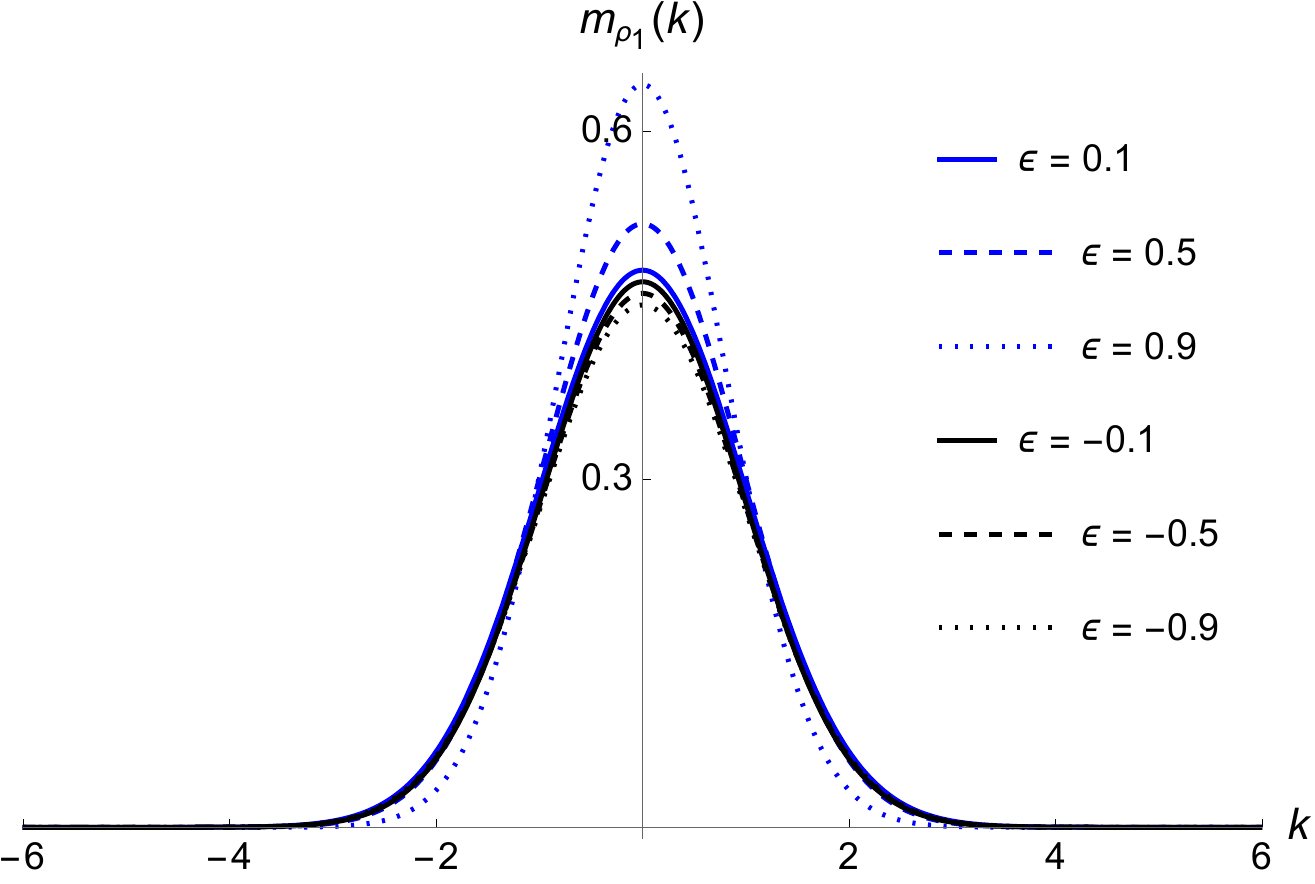}\qquad
\caption{Modal fraction for the topological sector $\protect\upphi_1(x)$, for  several values of $\upepsilon$.}
\label{fig8}\eec
\end{figure}
\begin{figure}[H]\bec
\includegraphics[scale=0.41]{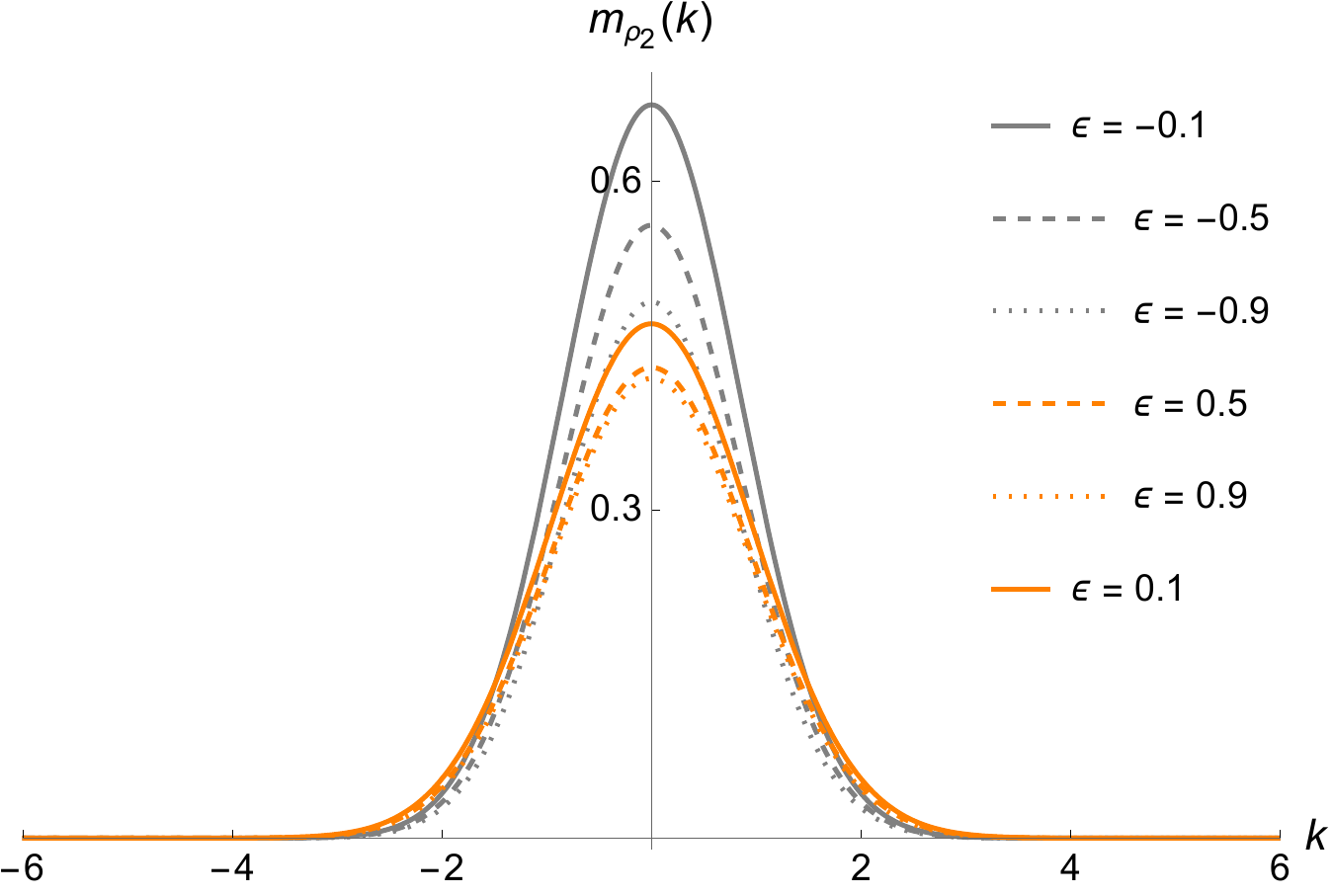}\qquad
\caption{Modal fraction for the topological sector $\protect\upphi_2(x)$, for  several values of $\upepsilon$.}\label{fig10}\eec
\end{figure}

\begin{figure}[H]
\bec
\includegraphics[scale=0.41]{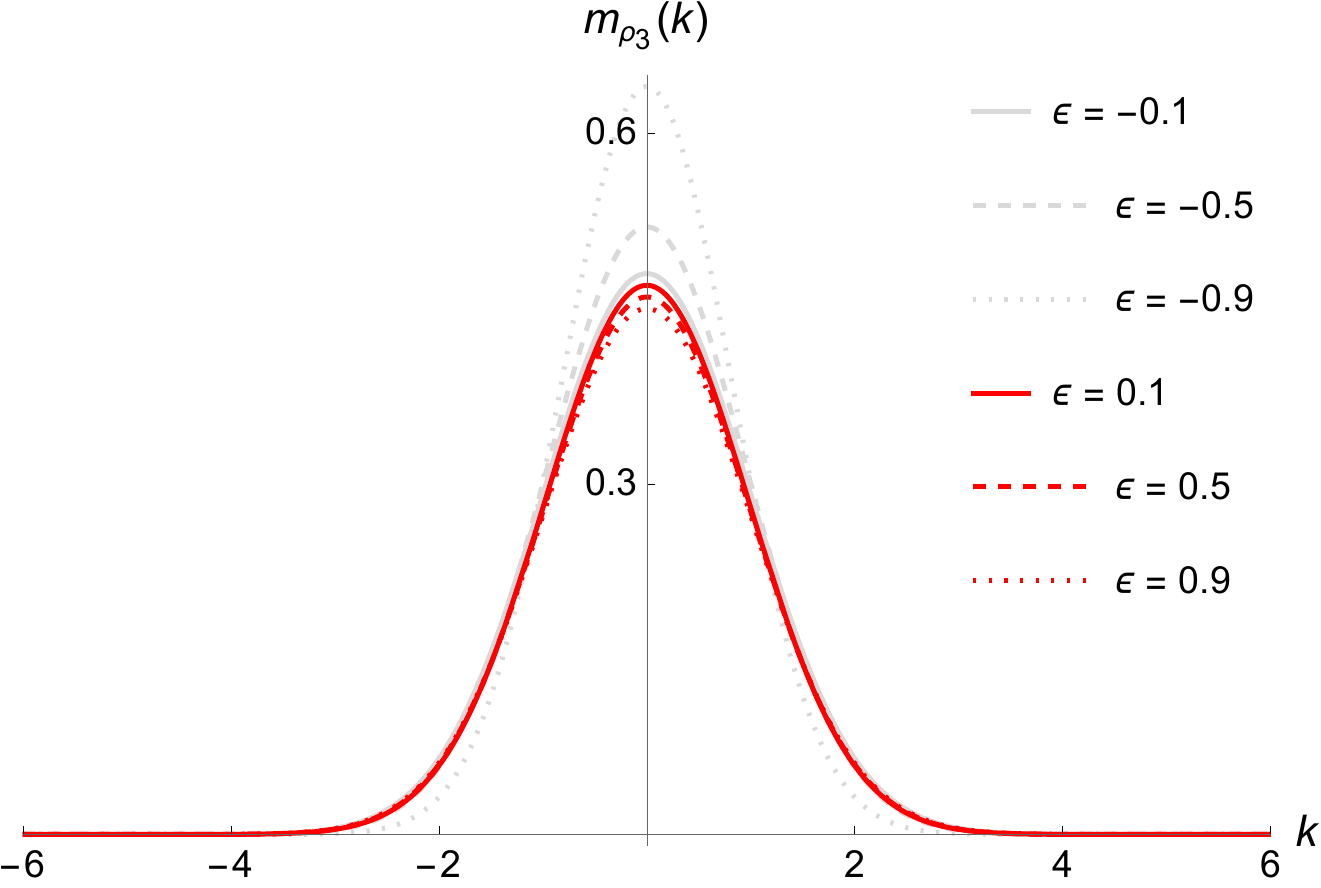}\qquad
\caption{Modal fraction for the topological sector $\protect\upphi_3(x)$, for  several values of $\upepsilon$.}\label{fig12}\eec
\end{figure}
For all topological sectors, the modal fraction peaks at $\mathsf{k}=0$, as expected. 
Fig. \ref{fig8} depicts the modal fraction of the first topological sector (\ref{eq1}), encoded in its expression for the energy density (\ref{ro1}). For $\upepsilon>0$ the absolute value of the peak of the modal fraction considerably varies as a function of $\upepsilon$, with the maximal variation of $\sim 48.1\%$ in the range here analyzed. When $\upepsilon<0$, the relative variation reduces to $\sim 3.8\%$. Now Fig. \ref{fig10} represents the modal fraction of the second topological sector (\ref{eq2}), encoded in its expression for the energy density (\ref{ro2}). This time, for $\upepsilon<0$ the absolute value of the peak of the modal fraction considerably varies as a function of $\upepsilon$, with the maximal variation of $\sim 36.7\%$ in the range analyzed. When $\upepsilon>0$, the relative variation reduces to $\sim 10.5\%$. Fig. \ref{fig12} analyzes the modal fraction of the third topological sector (\ref{eq3}), encoded in its expression for the energy density (\ref{rho123}). For $\upepsilon<0$ the absolute value of the peak of the modal fraction again considerably changes as a function of $\upepsilon$, with the maximal variation of $\sim 29.3\%$, in the range of $\upepsilon$ analyzed. When $\upepsilon>0$, the relative variation reduces to $\sim 2.6\%$.

\clt{Now, for the numerical computation of DCE by Eq. (\ref{ce1}), as we are dividing by the supremum, the DCC has units of ${\rm length}^{-1}$. Therefore, in Fig. \ref{dce1} this length is restored by considering a real interval $[-b,b]$, wherein the difference between the values of topological solutions  $\upphi_i$, $i=1,2,3$, respectively in Eqs. (\ref{eq1}) -- (\ref{eq3}), and their respective asymptotic values is greater than $10^{-10}$. Hence an effective length scale $\mathring{\ell}=|2b|$ can be inferred for each topological solution, meaning the effective size of the kinks out of their asymptotic values. For the solution $\upphi_1(x)$ in  (\ref{eq1}), $\mathring{\ell}_1\approxeq 44.97$ and for $\upphi_2(x)$ in  (\ref{eq2}) it follows that $\mathring{\ell}_2\approxeq 47.81$, whereas  for $\upphi_3(x)$ in  (\ref{eq3}) implies $\mathring{\ell}_3\approxeq 48.81$. These values are not significantly altered by the values of $\upepsilon$ and will be adopted respectively as the effective length scale.  }

Fig. \ref{dce1} depicts the DCE (\ref{ce1}), to each topological
sector.
\begin{center}\begin{figure}[H]\centering
\qquad\includegraphics[scale=0.66]{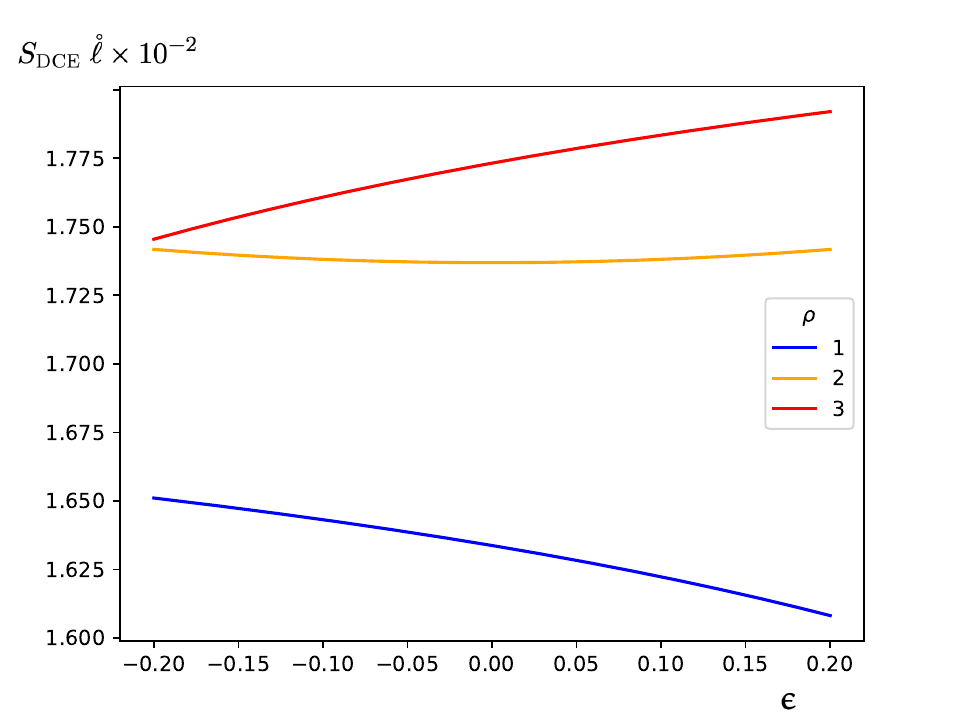}
\caption{DCE $\times \mathring{\ell}$ for the topological sectors $\protect\upphi%
_1(x)$, $\upphi_2(x)$, and $\upphi_3(x)$ as a function of the $\upepsilon$-deformation parameter.}\label{dce1}
\end{figure}
\end{center}
\clt{There is a value of minimal configurational entropy, with respect to $\upepsilon$, for the second topological sector \eqref{eq2}, with energy density \eqref{ro2}. Thus DCE points to the breaking of the degeneracy of the 3-sine-Gordon model, since the second topological sector \eqref{ro2} has a global minimum with DCE given by $S_{\scalebox{.55}{\textsc{DCE}}} = 3.656 \times 10^{-4}$ nat, for $\upepsilon\approxeq0$. For values $|\upepsilon|>0$ the DCE satisfies $d^2S_{\scalebox{.55}{\textsc{DCE}}}/d\upepsilon^2>0$. It also increases monotonically for $\upepsilon>0$, with $dS_{\scalebox{.55}{\textsc{DCE}}}/d\upepsilon>0$ and for $\upepsilon>0$, whereas $dS_{\scalebox{.55}{\textsc{DCE}}}/d\upepsilon<0$ for $\upepsilon>0$. For the first topological sector \eqref{eq1}, with energy density \eqref{ro1}, the DCE is a monotonically increasing function of $\upepsilon$ with no maxima with $dS_{\scalebox{.55}{\textsc{DCE}}}/d\upepsilon = 0$, including the infimum of the range of $\upepsilon$ analyzed. For the third topological sector \eqref{eq3}, the DCE is a monotonically decreasing function of $\upepsilon$, also with no minima with $dS_{\scalebox{.55}{\textsc{DCE}}}/d\upepsilon = 0$, including the supremum of $\upepsilon$ in the range analyzed. For both the first and third topological sectors, $d^2S_{\scalebox{.55}{\textsc{DCE}}}/d\upepsilon^2$ is negative. Besides, the first topological sector $\protect\upphi_1(x)$ presents higher configurational stability compared to both the second and third topological sectors. In its turn, the second topological sector $\protect\upphi%
_2(x)$ is more stable than $\protect\upphi_3(x)$, from the configurational entropic point of view. }

\clt{Now the DCC (\ref{dcc}) can be analyzed. It is illustrated in Fig. \ref{dcc1}, to each topological
sector.}
\begin{center}\begin{figure}[H]\centering
\qquad\includegraphics[scale=0.66]{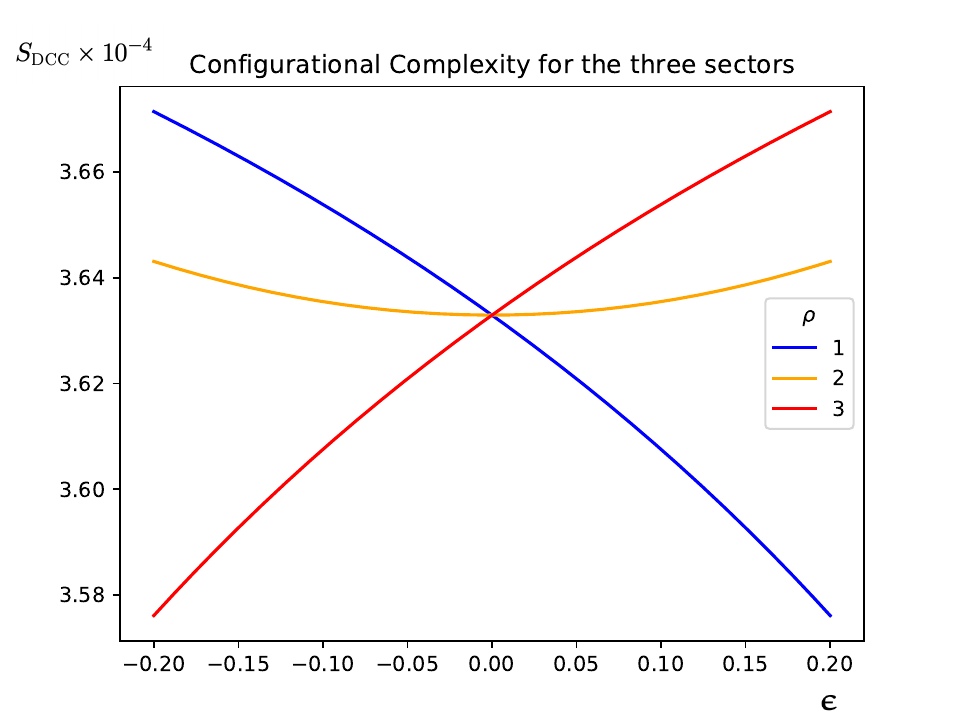}
\caption{DCC for the topological sectors $\protect\upphi%
_1(x)$, $\upphi_2(x)$, and $\upphi_3(x)$.}\label{dcc1}
\end{figure}
\end{center}
\clt{There is a value of minimal complexity with respect to $\upepsilon$ for the second topological sector \eqref{eq2}, with energy density \eqref{ro2}. Thus DCC can break the degeneracy of the 3-sine-Gordon model. In fact, the second topological sector \eqref{ro2} has a global minimum with DCC given by $S_{\scalebox{.55}{\textsc{DCC}}} = 3.634 \times 10^{-4}$ nat, for $\upepsilon\approxeq0$. For values $|\upepsilon|>0$ the DCC satisfies $d^2S_{\scalebox{.55}{\textsc{DCC}}}/d\upepsilon^2>0$. It also increases monotonically for $\upepsilon>0$, with $dS_{\scalebox{.55}{\textsc{DCC}}}/d\upepsilon>0$ and for $\upepsilon>0$, whereas $dS_{\scalebox{.55}{\textsc{DCC}}}/d\upepsilon<0$ for $\upepsilon>0$. For the first topological sector \eqref{eq1}, with energy density \eqref{ro1}, the DCC is a monotonically decreasing function of $\upepsilon$ with no maxima with $dS_{\scalebox{.55}{\textsc{DCC}}}/d\upepsilon = 0$, including the supremum corresponding to the first point of the range of $\upepsilon$ analyzed. For the third topological sector \eqref{eq3}, with energy density \eqref{rho123}, the DCC is a monotonically increasing function of $\upepsilon$, also with no minima with $dS_{\scalebox{.55}{\textsc{DCC}}}/d\upepsilon = 0$, including the infimum regarding the last point of the range of $\upepsilon$ analyzed. For both the first and third topological sectors, the inequality $d^2S_{\scalebox{.55}{\textsc{DCC}}}/d\upepsilon^2<0$ holds.
The degenerate (second) sector has the degeneracy broken by the DCC. }

\subsection{An additional approach to the 3-sine-Gordon model}
\label{ssec32}
Now the DCE of another deformation of the 3-sine-Gordon model, without a power expansion approach, will be derived. Refs. \cite{Epl1,Bazeia:2002xg} employed the deformation protocol to engender a new family of sine-Gordon-type models, including the 3-sine-Gordon. The general sine-Gordon-type models are regulated by two real parameters. The first one, $a$, controls both the location and the height of the scalar field maxima, whereas the second parameter, $b$, indexes the constituents in the family of sine-Gordon-type models, also managing the number of disconnected topological sectors \cite{Bazeia:2008tj}. Turning back to the results in Sec. \ref{sec2}, taking into account the $\upchi^4$ model with spontaneous symmetry breaking,
\beq
V(\upchi) = \frac12(1-\upchi^2)^2,
\eeq
Eq. \eqref{start} yields \cite{Bazeia:2011ff}
\be\label{equ8}
U(\upphi)=\frac12\frac{(1-f^2(\upphi))^2}{f^{2}_\upphi(\upphi)},
\ee
since $\upphi(x) = f^{-1}(\upchi(x))$. Eq. (\ref{equ8}) is invariant under the inverse mapping $f(\upphi)\mapsto 1/f(\upphi)$. Hence, as $f(\upphi)$ is the functional deforming the model, $1/f(\upphi)$ also deforms the model in the same way. This function can carry indexes $a$ and $b$ and is given by \cite{Epl1,Bazeia:2002xg}
\be\label{def11}
f_{ab}(\upphi)=\tan\{a \arctan[b\tan(\upphi)]\},
\ee
where $b\in(0,1)$ or $b\in(1,\infty)$, and $a\in\mathbb{N}$, 
 with potential
\beq\label{v12}
V(\upphi)&=&\frac{1}{2b^2a^2}\left\{2\cos^2[a \arctan(b\tan(\upphi))]-1\right\}^2\left[(1-b^2)\cos^2(\upphi)+b^2\right]^2.
\eeq
The parameter $a$ plays a prominent role in the generation of further families of models. For $a=1$ the double sine-Gordon model is acquired, which contains two topological sectors and for $a=2$ the triple sine-Gordon model is obtained, which contains three distinct topological sectors. The parameter $b$ controls the position of the minima and the height of the maxima. 
The topological solutions are given by, for a given integer $a$ \cite{Epl1,Bazeia:2002xg} 
\beq\label{alls}
\upphi^{m}_{k,a}&=&\arctan\left[\frac1b\tan\left(\frac1a\arctan\upchi(x)+k\frac{\pi}{a}\right)\right]\pm m\pi,
\een
where $k,m\in\mathbb{N}$, with \beq
\upchi(x)=\pm\tanh(x),\eeq giving rise to the defect and anti-defect solutions. 
For $a=2$, namely, for the 3-Gordon model, the superpotential reads 
\ben
\hspace*{-0.7cm}W(\upphi)&=&\frac{(1\!+\!b^2)(1\!-\!10b^2\!+\!b^4)}{2(1-b^2)^2}\upphi+\frac{(1\!+\!6b^2\!+\!b^4)}{4(1-b^2)} \sin(2\upphi)+\frac{8b^3}{(1-b^2)^2} \arctan[b\tan(\upphi)]\,.
\een
The energy densities of the three distinct topological sectors are provided respectively by \cite{Epl1,Bazeia:2002xg} 
\bes\ben
\hspace*{-0.8cm}\rho_1(x)&=&\frac{b^2{\rm sech}^2(2x)}{4\left
[(1\!-\!b^2)\cos^2[{\rm arctan}[\tanh(x)]]\!-\!1\right]^2},\label{r1}
\\
\hspace*{-0.8cm}\rho_2(x)&=&\frac{b^2{\rm sech}^2(2x)}{4\left[(1\!-\!b^2)\cos^2\left({\rm arctan}[\tanh(x)]\!+\!\frac{\pi}{4}\right)\!-\!1\right]^2},\label{r2}
\\
\hspace*{-0.8cm}\rho_3(x)&=&\frac{b^2{\rm sech}^2(2x)}{4\left[(1\!-\!b^2)\cos^2\left({\rm arctan}[\tanh(x)]\right)\!+\!b^2\right]^2}\,,\label{r3}
\een\ees

Using the protocol (\ref{ftrans}, \ref{modalf}, \ref{ce1}), the DCE can be computed. The plots in Figs. \ref{dce2a} -- \ref{dce2c}   show the DCE of the topological sectors $\protect\upphi_1(x)$, $\protect\upphi_2(x)$, and $\protect\upphi_3(x)$, with energy densities respectively given by Eqs. (\ref{r1}) -- (\ref{r3}), as a function of the parameter $b$. 
Contrary to the analytical Fourier transforms of the energy densities (\ref{ro1} -- \ref{rho123}) in the Appendix \ref{aa}, there is no analytical 
expressions for the energy densities (\ref{r1}) -- (\ref{r3}). 
Therefore, the DCE protocol (\ref{ftrans}, \ref{modalf}, \ref{ce1}) is
 numerically computed for this model. \clt{Analogously to the 
 numerical computation of DCE for the $\upepsilon$-deformed 
 model,   an effective length scale can be inferred for each topological 
 solution, corresponding to the domain in the $x$ coordinate where 
 the kinks do not coincide with their asymptotic values, up to $10^{-10}$.  
 For the solution $\upphi_1(x)$ in  (\ref{eq1}), $\mathring{\ell}_1\approxeq 18.92$ and for $\upphi_2(x)$ in  (\ref{eq2}) it follows 
 that $\mathring{\ell}_2\approxeq 21.13$, whereas  for $\upphi_3(x)$ 
 in  (\ref{eq3}) implies $\mathring{\ell}_3\approxeq 29.12$. These 
 values are not significantly altered by the values of $\upepsilon$ 
 and will be adopted respectively as the effective length scale.  }
\begin{figure}[H]\bec
\includegraphics[scale=0.56]{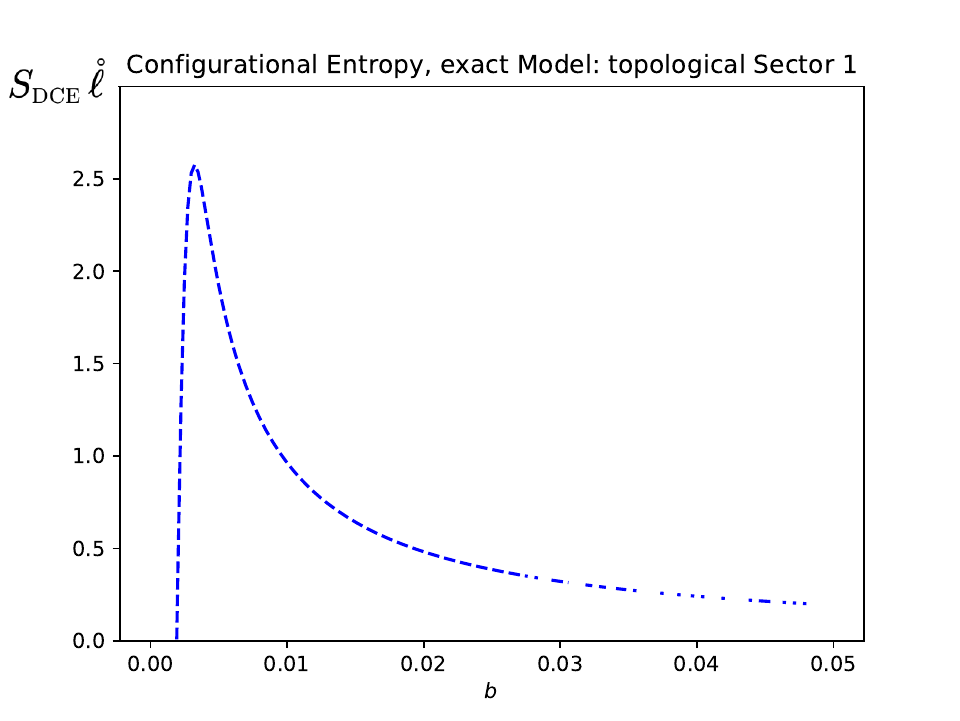} %
\caption{DCE $\times \mathring{\ell}$  of the topological sector $\protect\upphi_1(x)$ as a function of the parameter $b$.}\label{dce2a}\eec
\end{figure}
\noindent 

\begin{figure}[H]\bec
\includegraphics[scale=0.56]{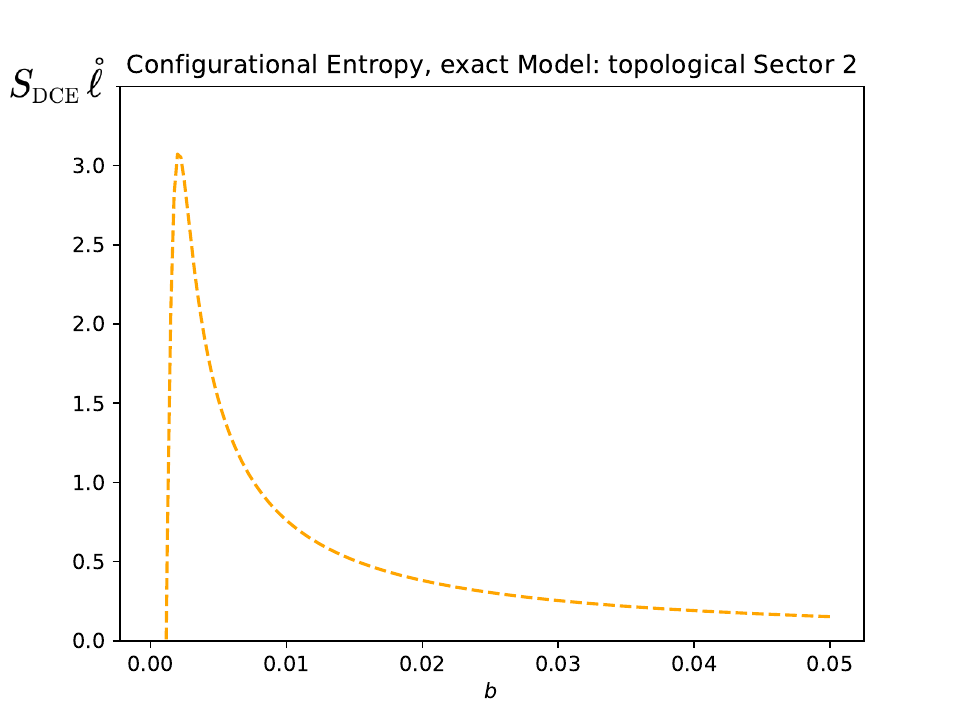} %
\caption{DCE $\times \mathring{\ell}$  of the topological sectors $\protect\upphi_2(x)$ as a function of the parameter $b$.}\label{dce2b}\eec
\end{figure}
\noindent 

\begin{figure}[H]\bec
\includegraphics[scale=0.56]{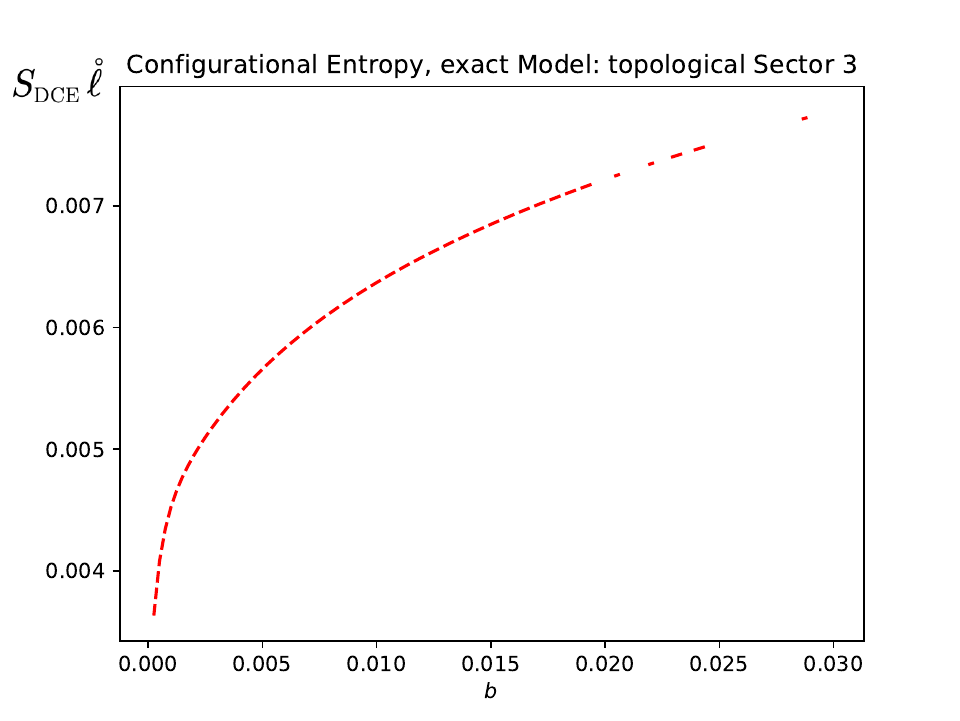}
\caption{DCE $\times \mathring{\ell}$  of the topological sectors $\protect\upphi_3(x)$ as a function of the parameter $b$.}\label{dce2c}\eec
\end{figure}
\noindent

Fig. \ref{dce2a} shows that the topological sector (\ref{r1}) has the DCE with a peak value $S_{\scalebox{.55}{\textsc{DCE}}} = 0.138$ nat, at $b = 0.0043$. In the interval $b\in(0,0.0043)$, the graphic of the DCE satisfies $d^2S_{\scalebox{.55}{\textsc{DCE}}}/db^2<0$ and $dS_{\scalebox{.55}{\textsc{DCE}}}/db>0$, whereas in the range $b\gtrsim 0.0043$ the DCE is a function of $b$ such that $d^2S_{\scalebox{.55}{\textsc{DCE}}}/db^2>0$ and $dS_{\scalebox{.55}{\textsc{DCE}}}/db<0$. Fig. \ref{dce2b}  illustrates the topological sector (\ref{r2}) with qualitatively analog profile of (\ref{r1}). However, it has the DCE with a peak value $S_{\scalebox{.55}{\textsc{DCE}}} = 0.147$ nat, at $b = 0.0022$. Regarding this plot, the DCE regarding (\ref{r2}) increases sharply and steeper than the case (\ref{r1}). Besides, in the range $b\in(0,0.0022)$, the graphic of the DCE satisfies $d^2S_{\scalebox{.55}{\textsc{DCE}}}/db^2<0$ and $dS_{\scalebox{.55}{\textsc{DCE}}}/db>0$, whereas when $b\gtrsim 0.0022$ the DCE satisfies $d^2S_{\scalebox{.55}{\textsc{DCE}}}/db^2>0$ and $dS_{\scalebox{.55}{\textsc{DCE}}}/db<0$. In this case, the peak has a higher DCE, attained at a lower value of the parameter $b$, when compared to the topological sector (\ref{r1}). Both cases (\ref{r1}) and (\ref{r2}) have the DCE asymptotically approaching zero, for $b\gtrsim 0.4$. We can conclude that 
for $b\in(0.4,1)$, the associated topological sectors with energy densities (\ref{r1}) and (\ref{r2}) are more stable, from the configurational point of view. The range $b\in(0.4,1)$ consists of the parameter space where the associated topological sector is more prevalent. For the topological sector (\ref{r2}), the neighborhood $0.0020\lesssim b \lesssim 0.0025$ represents the domain of higher configurational instability, whereas the interval $0.0040\lesssim b \lesssim 0.0046$ regards the domain of higher configurational instability of the topological sector (\ref{r1}). On the other hand, the DCE of the topological sector (\ref{r3}) \clt{in Fig. \ref{dce2c}}  has a completely distinct profile, with respect to the parameter $b$. It has no maxima, it is monotonically increasing with ${d^2 S_{\scalebox{.55}{\textsc{DCE}}}}/{db^2}<0$, for all values of $b$ in the range analyzed. The minimum of the DCE for the topological sector (\ref{r3}) occurs at $b\approxeq0$, for which $S_{\scalebox{.55}{\textsc{DCE}}} \approxeq 1.249$ nat. This is therefore the most stable configuration of the physical system. }

\clt{Now, the DCC is plotted in Fig. \ref{dcc2}.
\begin{figure}[H]
\includegraphics[scale=0.56]{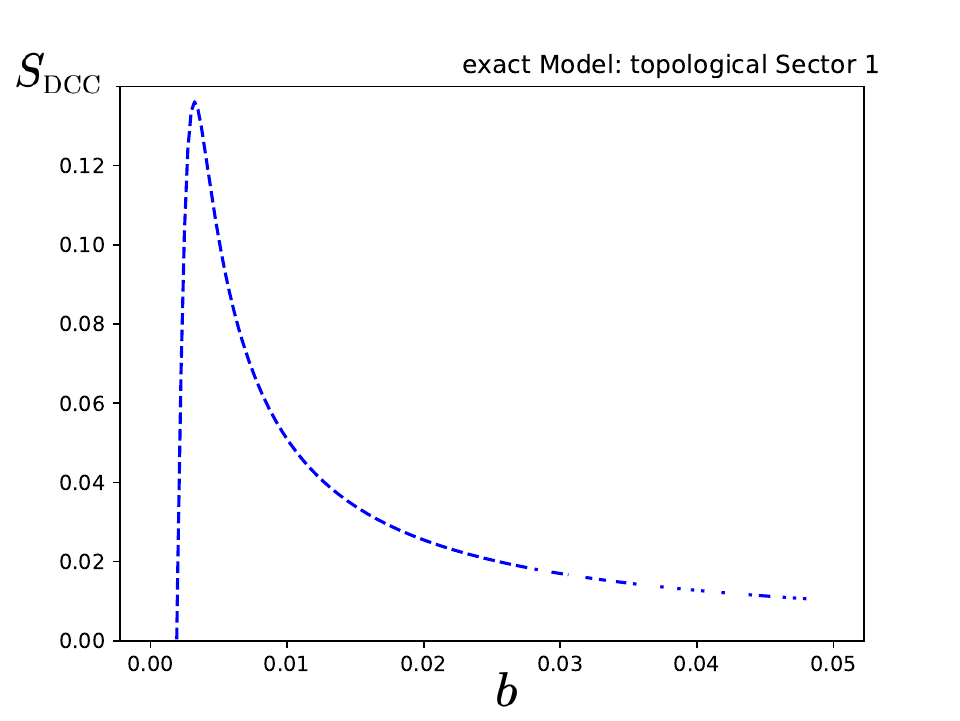} \!\!\!\!\!\!\!\!\!\!\!\!
\includegraphics[scale=0.56]{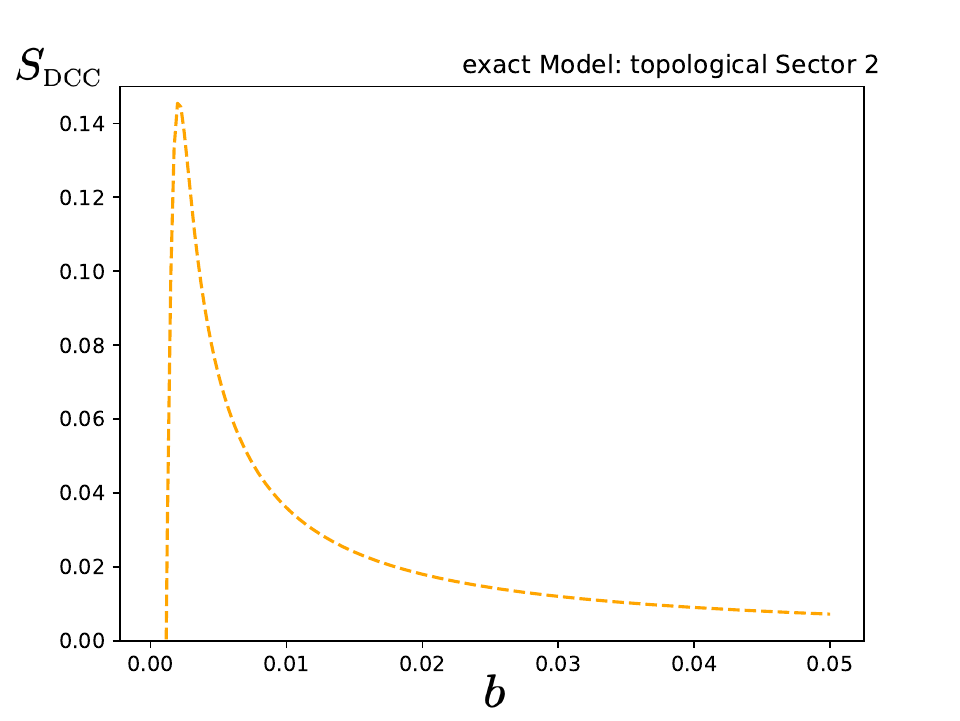} %
\caption{DCC of the topological sectors $\protect\upphi_1(x)$ (left) and $\protect\upphi_2(x)$ (right), as a function of the parameter $b$.}\label{dcc2}
\end{figure}
\begin{figure}[H]\bec
\includegraphics[scale=0.56]{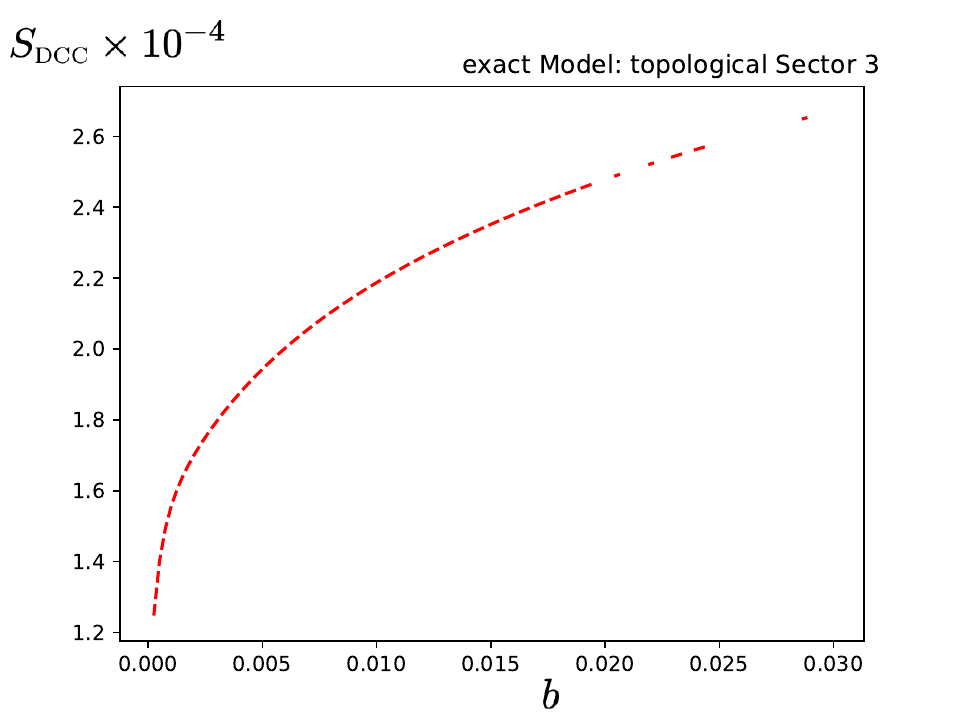} %
\caption{DCC of the topological sector $\protect\upphi_3(x)$, as a function of the parameter $b$.}\label{dcc21}\eec
\end{figure}}
\noindent\clt{The plots of the DCC in Figs. \ref{dcc2} and \ref{dcc21} carry a similar profile to the DCE in Figs. \ref{dce2a} -- \ref{dce2c}.}

\clt{Finally, the DCC can be analyzed as a function of the energies of the kinks.  Since the energies (\ref{e123}) of the first and the third topological sector are $\upepsilon$-(linearly)-dependent, the plots are depicted in Fig. \ref{dcc2ee}. The plot of the DCC as a function of the energy $E_2$ of the second topological sector is omitted, since $E_2$ (\ref{e123}) is constant}.
\begin{figure}[H]
\includegraphics[scale=0.56]{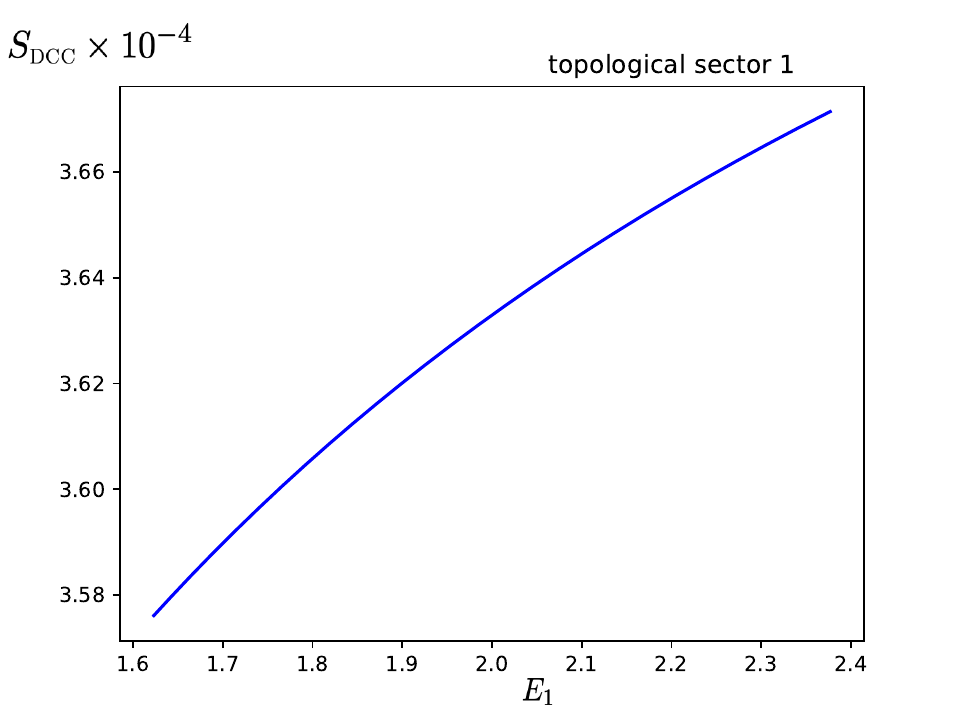} \!\!\!\!\!\!\!\!\!%
\includegraphics[scale=0.56]{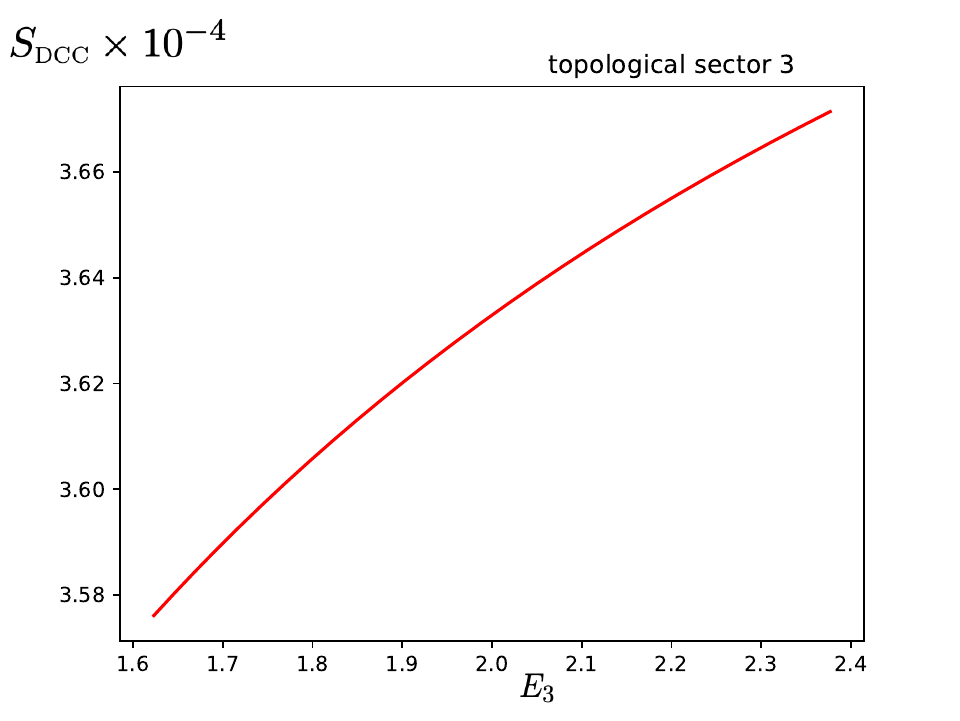} %
\caption{DCC of the topological sectors $\protect\upphi_1(x)$ (left) and $\protect\upphi_3(x)$ (right), as a function of the energy  $E$ of the kink.}\label{dcc2ee}
\end{figure}}

\section{concluding remarks and perspectives}
\label{sec5}

The protocol of deforming topological defects was here engendered to study two approaches of the 3-sine-Gordon model whose underlying DCE was computed and discussed.
Contrary to the approach in Sec. \ref{sec2}, where the deformation function could be first-order expanded with respect to the deformation parameter $\upepsilon$ as in Eq. (\ref{def1}), the analytical deformation function (\ref{def11}) in Sec. \ref{sec3} does not involve a power expansion. The results involving the underlying DCE \clt{and the DCC} analysis, for the two deformed models, are complementary. The deformations of the 3-sine-Gordon model can be employed for practical applications in a diversity of scenarios, in particular in the braneworld scenario with one extra dimension, being the brane stabilized by a real scalar field with the best choice of parameters driven by the \clt{DCE and the DCC}, as detailedly reported in the final discussion of Subsecs. \ref{ssec3} and \ref{ssec32}, respectively for each different deformation procedure generating the 3-sine-Gordon model.  
Ref. \cite{epjc} addressed deformed 2-sine-Gordon models, with applications in gravity localization on thick braneworlds. Here, the results can be used to study gravity localization with deformed 3-sine-Gordon field configurations. The results heretofore derived in this work, using the DCE, can hence drive the best choice of the range of parameters for studying thick branes and field localization on them. \clt{As the approach of using deformations has been successfully applied to the study of braneworld scenarios, especially in what concerns the localization of bosonic and fermionic fields onto the brane, one can investigate the relevant results in Refs. \cite{Bazeia:2017nlo,Bazeia:2013euc,Bazeia:2014xqa,Bazeia:2014dea,Afonso:2006gi,Herrera-Aguilar:2014oua}, including asymmetric braneworlds in Refs. \cite{Bazeia:2015fca}, using the DCE. Finally, higher-dimensional deformation techniques \cite{Afonso:2006ws,Bazeia:2013nma} may be approached by DCE techniques.}
\acknowledgments
RdR~is grateful to FAPESP (Grant No. 2021/01089-1 and No. 2022/01734-7) and the National Council for Scientific and Technological Development -- CNPq (Grant No. 303390/2019-0), for partial financial support.
AHA has benefited from the grants CONACYT No. A1-S-38041 and VIEP-BUAP No. 122. 
\appendix
\section{Fourier transforms of the energy density for each topological sector (\ref{ro1}) -- (\ref{rho123})}
\label{aa}
{\small{\begin{widetext}
\begin{eqnarray}
&&\left. \rho_1(\mathsf{k})={\mathsf{k}} \sqrt{\frac{\pi }{2}}\csc\!{\rm h}\left( \frac{%
{\mathsf{k}} \pi }{2}\right) -\frac{\upepsilon }{6\sqrt{\pi }}\left\{ \frac{i}{%
{\mathsf{k}} +2i}\left[ (9+3i{\mathsf{k}} +2{\mathsf{k}} ^{2})_{2}F_{1}\left( -\frac{1}{2},1-%
\frac{i{\mathsf{k}} }{2};2-\frac{i{\mathsf{k}} }{2};-1\right) \right. \right. \right.  
\notag \\
&&\left. \left.  \qquad\qquad\qquad\qquad\qquad\qquad\qquad\qquad-(7+2i{\mathsf{k}} +4{\mathsf{k}} ^{2})_{2}F_{1}\left( \frac{1}{2},1-%
\frac{i{\mathsf{k}} }{2};2-\frac{i{\mathsf{k}} }{2};-1\right) \right] \right. 
\notag \\
&&\qquad\qquad\qquad\qquad\qquad\qquad\qquad\qquad \left. +\frac{1}{{\mathsf{k}}
-2i}\left[ -i(9-3i{\mathsf{k}} +2{\mathsf{k}} ^{2})_{2}F_{1}\left( -\frac{1}{2},1+\frac{%
i{\mathsf{k}} }{2};2+\frac{i{\mathsf{k}} }{2};-1\right) \right. \right.  \notag \\
&&\left. \left.\qquad\qquad\qquad\qquad\qquad\qquad\qquad\qquad +(7i+2{\mathsf{k}} +4i{\mathsf{k}} ^{2})_{2}F_{1}\left( \frac{1}{2},1+%
\frac{i{\mathsf{k}} }{2};2+\frac{i{\mathsf{k}} }{2};-1\right) \right]  \right.   \notag \\
&& \left.\qquad\qquad\qquad\qquad\qquad\qquad\qquad\qquad +\frac{1}{{\mathsf{k}}
-3i}\left[ -i(4-7i{\mathsf{k}} +2{\mathsf{k}} ^{2})_{2}F_{1}\left( -\frac{1}{2},\frac{3}{%
2}+\frac{i{\mathsf{k}} }{2};\frac{5}{2}+\frac{i{\mathsf{k}} }{2};-1\right) \right.
\right.  \notag \\
&&\left. \left.\qquad\qquad\qquad\qquad\qquad\qquad\qquad\qquad +(i+10{\mathsf{k}} +4i{\mathsf{k}} ^{2})_{2}F_{1}\left( \frac{1}{2},%
\frac{3}{2}+\frac{i{\mathsf{k}} }{2};\frac{5}{2}+\frac{i{\mathsf{k}} }{2};-1\right) %
\right]  \right.
 \notag \\
&& \left.\qquad\qquad\qquad\qquad\qquad\qquad\qquad\qquad +\frac{1}{{\mathsf{k}} +3i}\left[ (4+7i{\mathsf{k}} +2{\mathsf{k}}
^{2})_{2}F_{1}\left( -\frac{1}{2},\frac{3}{2}-\frac{i{\mathsf{k}} }{2};\frac{5}{2}-%
\frac{i{\mathsf{k}} }{2};-1\right) \right. \right.  \notag \\
&&\left. \left.\qquad\qquad\qquad\qquad\qquad\qquad\qquad\qquad  -(1+10i{\mathsf{k}} +4{\mathsf{k}} ^{2})_{2}F_{1}\left( \frac{1}{2},-%
\frac{3}{2}+\frac{i{\mathsf{k}} }{2};\frac{5}{2}-\frac{i{\mathsf{k}} }{2};-1\right) %
\right] \right\} .
\end{eqnarray}
\end{widetext}}}\noindent 
where the above expressions $_{2}\emph{F}_{1}(A,\;B;\;C;\;D)$ stand for the
well-known hypergeometric functions. Moreover, for the sectors of the second
kind, we have
{\small{\begin{widetext} 
\begin{eqnarray}
&&\left. \rho_2(\mathsf{k})={\mathsf{k}} \sqrt{\frac{\pi }{2}}\csc\!{\rm h}\left( \frac{%
{\mathsf{k}} \pi }{2}\right) +\frac{\upepsilon }{\sqrt{2\pi }}\left\{ -\frac{%
i\upepsilon }{8({\mathsf{k}} +i)}\left[ G_{1}\left(1\!-\!i{\mathsf{k}};-\frac{1}{2},\frac{1}{%
2};2\!-\!i{\mathsf{k}},-i,i\right) \right. \right. \right.  \notag \\
&&\left. \qquad\quad\quad\qquad\qquad\qquad\qquad\qquad +e^{i\pi/4}G_{1}\!\left( 1\!-\!i{\mathsf{k}} ;-\frac{1}{2},\frac{3}{2};2\!-\!i{\mathsf{k}}
,-i,i\right)\! -\!2e^{-i\pi/4}G_{1}\left( 1\!-\!i{\mathsf{k}} ;-\frac{1}{2},\frac{5}{2}%
;2\!-\!i{\mathsf{k}} ,-i,i\right)\!  \right.  \notag \\
&&\left. \qquad\quad\quad\qquad\qquad\qquad\qquad\qquad +G_{1}\left( 1\!-\!i{\mathsf{k}} ;\frac{1}{2},-\frac{1}{2}%
;2\!-\!i{\mathsf{k}} ,-i,i\right)  +e^{-i\pi/4}G_{1}\left( 1\!-\!i{\mathsf{k}} ;\frac{3}{2},-\frac{1}{2}%
;2\!-\!i{\mathsf{k}} ,-i,i\right)\! \right.  \notag \\
&&\left. \left. \qquad\quad\quad\qquad\qquad\qquad\qquad\qquad -2e^{i\pi/4}G_{1}\left( 1\!-\!i{\mathsf{k}} ;\frac{5}{2},-\frac{1}{2%
};2\!-\!i{\mathsf{k}} ,-i,i\right) \right] \right.  \notag \\
&&\left.  \qquad\quad\quad\qquad\qquad\qquad\qquad\qquad -\frac{1}{3\sqrt{2}}\left[ \frac{1}{{\mathsf{k}} -2i }%
\left((9i+3{\mathsf{k}}-2i{\mathsf{k}} ^{2})_{2}F_{1}\left(\!-\frac{1}{2},1\!+\!\frac{%
i{\mathsf{k}} }{2};2\!+\!\frac{i{\mathsf{k}} }{2};-1\right) \right. \right. \right.  \notag \\
&&\left. \left.\left. \qquad\quad\quad\qquad\qquad\qquad\qquad\qquad+(7i+2{\mathsf{k}} +4i{\mathsf{k}}
^{2})_{2}F_{1}\left(\frac{1}{2},1\!+\!\frac{i{\mathsf{k}} }{2};2+\frac{i{\mathsf{k}} }{2}%
;-1\right) \right) \right. \right.  \notag \\
&&\left. \left. \qquad\quad\quad\qquad\qquad\qquad\qquad\qquad+\frac{1}{{\mathsf{k}}\!-\!3i}\left((4i+7{\mathsf{k}}-2i{\mathsf{k}}
^{2})_{2}F_{1}\left(-\frac{1}{2},\frac{3}{2}\!+\!\frac{i{\mathsf{k}} }{2};\frac{5}{2}\!+\!%
\frac{i{\mathsf{k}} }{2};-1\right)\right. \right.\right.  \notag \\&&\left.\left. \left. \qquad\quad\quad\qquad\qquad\qquad\qquad\qquad +(i\!+\!10{\mathsf{k}} \!+\!4i{\mathsf{k}} ^{2})_{2}F_{1}\left( 
\frac{1}{2},\frac{3}{2}\!+\!\frac{i{\mathsf{k}} }{2};\frac{5}{2}\!+\!\frac{i{\mathsf{k}} }{2}%
;-1\right) \right) \right] \right\} ,
\end{eqnarray}
\end{widetext}}}
\noindent where $\emph{G}_{1}(A,\;B;\;C;\;D)$ denotes the Appell hypergeometric
function. Finally,  the analysis of the third sector yields 
{\small{\begin{widetext}
\begin{eqnarray}
&&\left. \rho_3(\mathsf{k})={\mathsf{k}} \sqrt{\frac{\pi }{2}}\csc\!{\rm h}\left( \frac{%
{\mathsf{k}} \pi }{2}\right) +\frac{\upepsilon }{6\sqrt{\pi }}\left\{ \frac{i}{%
{\mathsf{k}} +2i}\left[ (9+3i{\mathsf{k}} +2{\mathsf{k}} ^{2})_{2}F_{1}\left( -\frac{1}{2},1-%
\frac{i{\mathsf{k}} }{2};2-\frac{i{\mathsf{k}} }{2};-1\right) \right. \right. \right.  
\notag \\
&&\left. \left. \qquad\quad\quad\qquad\qquad\qquad\qquad\qquad  -(7+2i{\mathsf{k}} +4{\mathsf{k}} ^{2})_{2}F_{1}\left( \frac{1}{2},1-%
\frac{i{\mathsf{k}} }{2};2-\frac{i{\mathsf{k}} }{2};-1\right) \right] +\right.\notag \\
&&\left. \qquad\quad\quad\qquad\qquad\qquad\qquad\qquad +\frac{1}{{\mathsf{k}}
-2i}\left[ -i(9-3i{\mathsf{k}} +2{\mathsf{k}} ^{2})_{2}F_{1}\left( -\frac{1}{2},1+\frac{%
i{\mathsf{k}} }{2};2+\frac{i{\mathsf{k}} }{2};-1\right) \right. \right.  \notag \\
&&\left. \left.\qquad\quad\quad\qquad\qquad\qquad\qquad\qquad +(7i+2{\mathsf{k}} +4i{\mathsf{k}} ^{2})_{2}F_{1}\left( \frac{1}{2},1+%
\frac{i{\mathsf{k}} }{2};2+\frac{i{\mathsf{k}} }{2};-1\right) \right] \right.   \notag \\
&&\left. \qquad\quad\quad\qquad\qquad\qquad\qquad\qquad  +\frac{1}{{\mathsf{k}}
-3i}\left[ -i(4-7i{\mathsf{k}} +2{\mathsf{k}} ^{2})_{2}F_{1}\left( -\frac{1}{2},\frac{3}{%
2}+\frac{i{\mathsf{k}} }{2};\frac{5}{2}+\frac{i{\mathsf{k}} }{2};-1\right) \right.
\right.  \notag \\
&&\left. \left. \qquad\quad\quad\qquad\qquad\qquad\qquad\qquad  +(i+10{\mathsf{k}} +4i{\mathsf{k}} ^{2})_{2}F_{1}\left( \frac{1}{2},%
\frac{3}{2}+\frac{i{\mathsf{k}} }{2};\frac{5}{2}+\frac{i{\mathsf{k}} }{2};-1\right) %
\right] \right.   \notag \\
&&\left. \qquad\quad\quad\qquad\qquad\qquad\qquad\qquad  +\frac{1}{{\mathsf{k}} +3i}\left[ (4+7i{\mathsf{k}} +2{\mathsf{k}}
^{2})_{2}F_{1}\left( -\frac{1}{2},\frac{3}{2}-\frac{i{\mathsf{k}} }{2};\frac{5}{2}-%
\frac{i{\mathsf{k}} }{2};-1\right) \right. \right.  \notag \\
&&\left. \left. \qquad\quad\quad\qquad\qquad\qquad\qquad\qquad  -(1+10i{\mathsf{k}} +4{\mathsf{k}} ^{2})_{2}F_{1}\left( \frac{1}{2},-%
\frac{3}{2}+\frac{i{\mathsf{k}} }{2};\frac{5}{2}-\frac{i{\mathsf{k}} }{2};-1\right) %
\right] \right\} .
\end{eqnarray}
\end{widetext}}}

\end{document}